UNIVERSITÄT KOBLENZ · LANDAU

Fachbereich 4: Informatik

gesis Leibniz Institute for the Social Sciences

WeST People and Knowledge Networks

# Measuring Gender Inequalities of German Professions on Wikipedia

## Master Thesis

A thesis submitted for the
Master of Science in Web Science

Submitted by

Olga Zagovora
Mat.-Nr. 214100694
zagovora@uni-koblenz.de

First supervisor: JProf. Dr. Claudia Wagner
(Institute for Web Science and Technologies,
University of Koblenz-Landau;
Computational Social Science Department,
GESIS - Leibniz Institute for the Social Sciences)

Second supervisor: Dr. Fabian Flöck
(Computational Social Science Department,
GESIS - Leibniz Institute for the Social Sciences)

Koblenz, October 2016

## Erklärung

Ich versichere, dass ich die vorliegende Arbeit selbständig verfasst und keine anderen als die angegebenen Quellen und Hilfsmittel benutzt habe.

| | Ja | Nein |
|---|---|---|
| Mit der Einstellung dieser Arbeit in die Bibliothek bin ich einverstanden. | ☐ | ☐ |
| Der Veröffentlichung dieser Arbeit im Internet stimme ich zu. | ☐ | ☐ |
| Der Text dieser Arbeit ist unter einer Creative Commons Lizenz verfügbar. | ☐ | ☐ |
| Der Quellcode ist unter einer Creative Commons Lizenz verfügbar. | ☐ | ☐ |
| Die erhobenen Daten sind unter einer Creative Commons Lizenz verfügbar | ☐ | ☐ |

. . . . . . . . . . . . . . . . . . . . . . . . . . . . . . . . . . . . . . . . . . . . . . . . . . . . . . . . . . . . . . . . .

(Ort, Datum) (Unterschrift)

## Abstract

Wikipedia is a community-created online encyclopedia; arguably, it is the most popular and largest knowledge resource on the Internet. Thus, reliability and neutrality are of high importance for Wikipedia. Previous research [3] reveals gender bias in Google search results for many professions and occupations. Also, Wikipedia was criticized for existing gender bias in biographies [4] and gender gap in the editor community [5, 6]. Thus, one could expect that gender bias related to professions and occupations may be present in Wikipedia. The term gender bias is used here in the sense of conscious or unconscious favoritism towards one gender over another [47] with respect to professions and occupations. The objective of this work is to identify and assess gender bias. To this end, the German Wikipedia articles about professions and occupations were analyzed on three dimensions: redirections, images, and people mentioned in the articles. This work provides evidence for systematic overrepresentation of men in all three dimensions; female bias is only present for a few professions.

## Acknowledgement

I would like to thank my supervisors Claudia Wagner and Fabian Flöck for their support and the great opportunities they have given to me over the last years. I would like to express my gratitude for their feedback throughout this work as they introduced me to scientific thinking.

Additionally, I would like to thank Markus Strohmaier, Computational Social Science Department at GESIS - Leibnitz Institute for the Social Sciences, for financially supporting this research. I would also like to thank the whole Data Science team, and namely Christoph Carl Kling, Lisa Posch, Fariba Karimi, Philipp Singer, and Florian Lemmerich, who inspired me and provided me with access to data sets.

Finally, I express my very profound gratitude to my loved ones, my parents and my partner for continuous encouragement throughout these years of study and providing me with unfailing support through the process of researching and writing this thesis. This accomplishment would not have been possible without them.

## Contents

## List of Figures

# 1. Introduction

## 1.1. Motivation

Wikipedia is an online encyclopedia that is ranked among the six most popular websites on the Internet in August 2016 [2]. According to WikiStats project[1], Wikipedia has 41.2 million articles [10] and had more than 15 billion page views in August 2016 [1]. Thus, Wikipedia is a popular source of information on the World Wide Web.

Online resources like Wikipedia may influence people's behaviors and perceptions about the world. The information people get from online media may affect their interpretations and understanding of the surrounding world. Biased information from media even can build new or endorse existing stereotypes of individuals [7]. It may influence a person's behavior towards other individuals. It may also influence a person's assessment of her or his career opportunities.

One of the most well-known biases is the so-called gender bias. Even though, Germany is in top 20 of the most gender-equal countries (countries with the best equality opportunities for women) according to the Global Gender Gap Index [8], gender disparities are still present. For example, the choices of professions remain gender-segregated due to the gender stereotypes of many professions[2]. Thus, many careers are gender-unbalanced which ultimately results in unequal compensations for women and men [9].

Recent studies [3, 4, 11] reveal the unequal representation of women and the way women are portrayed in community-created encyclopedia and image search results. Kay et al. [3] studied if and how gender ratios in image search results affects people's perceptions of the actual gender proportions in different professions. They found that for manipulated image search results people's perception slightly shifts towards the direction of the manipulated proportions. The results are also supported by cultivation theory [20], which suggests that long-term exposure to biased information might shift individuals' perceptions over time.

While representation, coverage, lexical, and visibility bias [4, 11, 12, 13] on Wikipedia received a lot of attention by the academic community, several aspects related to gender bias remain unexplored: people on images, people mentioned in articles, and gender inclusiveness in the use of job titles. This study aims to address this gap by studying Wikipedia articles along three dimensions: gender inclusiveness in job titles and corresponding redirection analysis, images analysis, and mentioned people analysis.

---

[1] https://stats.wikimedia.org/

[2] Further, we use the word professions in the thesis to refer to both professions and occupations.

To illustrate the overall issue of gender bias, let us look at some Wikipedia data. As a data point, consider the article “Journalist” on the German Wikipedia. Throughout the article, there are many images depicting men and only two images depicting women (Figure 1). Also, fewer women than men are mentioned in the text. A natural implicit or explicit conclusion could be that most journalists are male individuals.

This example illustrates that gender bias in Wikipedia may be an important problem and this work sets out to develop a computational approach to quantify such bias and make it transparent.

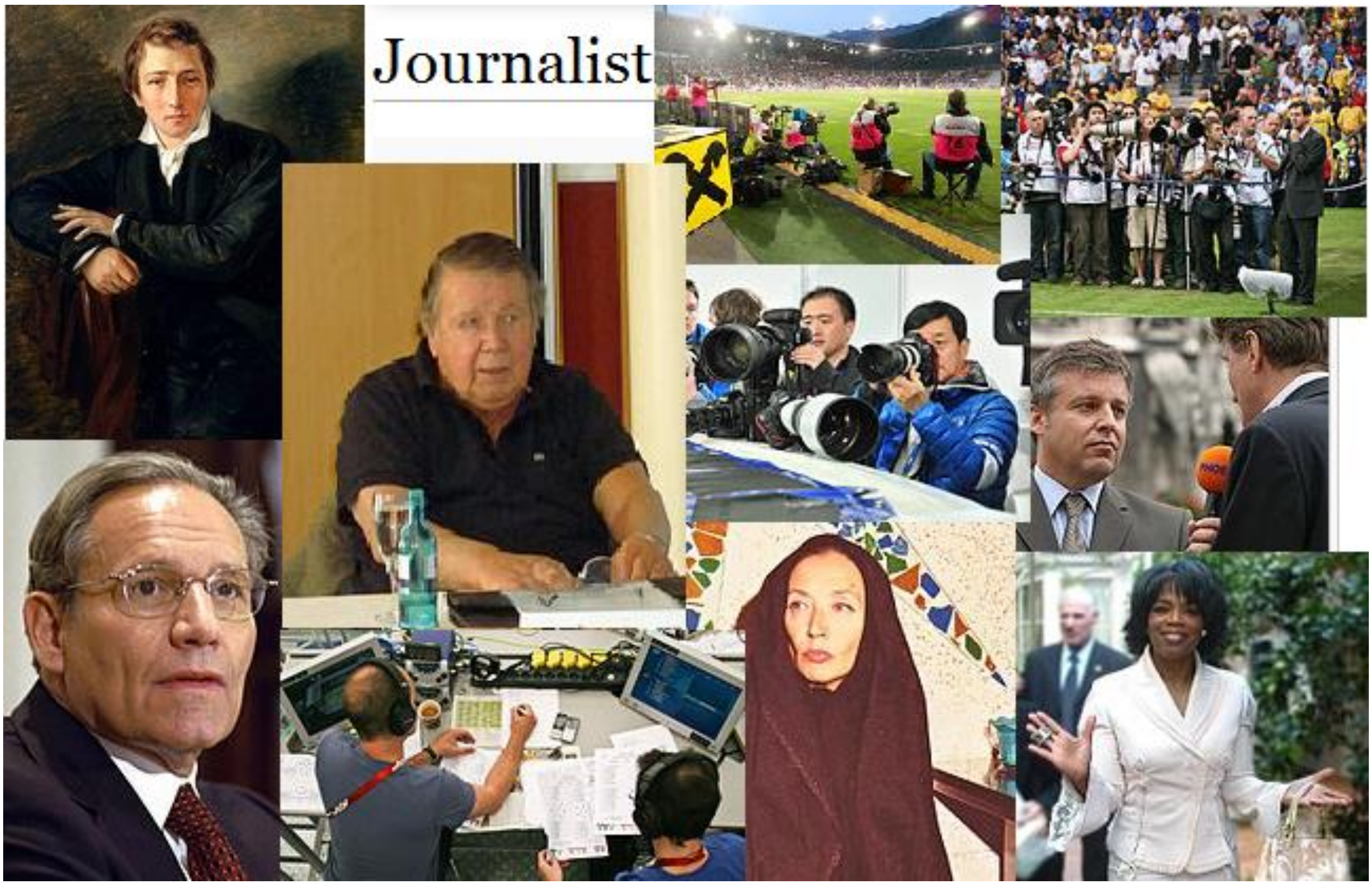

**Figure 1:** Collage of images obtained from the German Wikipedia article “Journalist”.

*Images from left upper corner:*
Heinrich Heine
Fotografen beim Fußball © Ralf Roletschek /CC-BY-SA-3.0/ http://creativecommons.org/licenses/by-sa/3.0
Fotojournalisten bei der Fußball-Europameisterschaft 2008 © Arne Müseler / arne-mueseler.de / CC-BY-SA-3.0 / https://creativecommons.org/licenses/by-sa/3.0/de/deed.de
Lothar Loewe bei einem Vortrag im Juli 2009 © Bücherhexe /CC-BY-SA-3.0 / http://creativecommons.org/licenses/by-sa/3.0
1. Jugendolympiade 2012 Innsbruck © Ralf Roletschek / CC-BY-SA-3.0 at / http://creativecommons.org/licenses/by-sa/3.0/at/deed.en
Reporter Heinz abel (PHOENIX) im Gespräch mit Peter Fahrenholz © André Zahn/ CC-BY-SA-2.0 de / http://creativecommons.org/licenses/by-sa/2.0/de/deed.en
Bob Woodward, assistant managing editor © Jim Wallace (Smithsonian Institution) / CC-BY-2.0 / http://creativecommons.org/licenses/by/2.0
Journalisten bei der Fußball-Europameisterschaft 2008 © Arne Müseler / arne-mueseler.de / CC-BY-SA-3.0 / https://creativecommons.org/licenses/by-sa/3.0/de/deed.de
Oriana Fallaci in Tehran 1979
Oprah Winfrey at the Hotel Bel Air in Los Angeles © Alan Light / CC-BY-2.0 / http://creativecommons.org/licenses/by/2.0

## 1.2. Approach and Research Question

To assess whether and to what extent the German Wikipedia is gender-biased with respect to professions, articles about professions are analyzed along three dimensions: gender inclusiveness in job titles and corresponding redirection analysis, images analysis, and mentioned people analysis. Initially, job titles were matched to Wikipedia articles using Levenshtein distance and ratio. Next, we checked whether existing articles have male, female, or neutral titles. We also checked if redirections occurred when particular profession names were requested.

Redirection analysis determines prevalence of male, female, or neutral job titles as Wikipedia article titles. To analyze Wikipedia articles along the dimension of images, all images from articles about professions were retrieved. Then genders of people on images were identified. Next, distributions of genders in the profession articles were compared. In this manner, gender inequalities in the use of images on Wikipedia can be revealed. The dimension of mentioned people reveals which gender is mentioned more often in the articles. To this end, people names from the text of articles were mined and their genders, according to their first names, were identified.

For all three dimensions, the results were compared with data from the labor market statistics of professions in Germany. The data includes numbers of men and women employed per profession.

The **objective** of this work is to identify and assess gender bias related to professions in Wikipedia articles. More precisely, we focus on the German Wikipedia and explore different dimensions of bias.

The main research questions of this thesis are the following:

1. *How can we measure gender bias related to professions in Wikipedia articles?*
2. *To what extent is gender bias present in articles about professions?*

## 1.3. Contributions and Findings.

The contributions of this work are as follows:

1. This study presents a computational approach to gender bias assessment along several dimensions using Wikipedia articles of profession domain. The proposed approach may be applicable to other Wikipedia language editions.
2. The empirical findings of this research are the following:
    - *Systematic overrepresentation of men* and *underrepresentation of women*: The results of analysis along three dimensions exhibit overrepresentation of men and underrepresentation of women. Analysis of article titles and redirections on Wikipedia reveals that most

professions are represented only via an article with a male title of profession. Moreover, most encountered redirections are from articles with female to male title. Along the images dimension, more men than women were depicted in most articles of the following groups: professions with female and male majority in the labor market. Overall, almost four times more images depicting men than women were observed, whereas the German labor market statistics exhibits 62.1% men and 37.9% women on average (if one considers professions which have Wikipedia articles). The results of the mentioned people analysis also reveal an underrepresentation of women, such that 83% mentioned men and only 17% women were observed on average in the profession articles.

- *Female bias for particular professions:* Results of analysis along three dimensions exhibit female bias for a few articles. The analysis of images reveals prevalence of images depicting women in the following groups: articles with female titles; professions with articles with female titles only. For example, professions midwife (German: “Hebamme”) and hostess (German: “Hostess”) are represented only with images depicting women. The analysis of mentioned people reveals that 3.1% of articles have female bias (i.e., articles that mention more than 50% women). If only people who were born after 1960 are considered, then 22% of articles have female bias.

The data and code are available online: https://github.com/gesiscss/Wikipedia-Language-Olga-master/ .

## 1.4. Overview

The remaining part of the thesis is structured as follows. In Chapter 2, related work regarding gender bias on Wikipedia and traditional media as well as methods for measuring gender bias is discussed. In Chapter 3, datasets and data collection methods are described. In Chapter 4, the research method is described. In Chapter 5, results are presented. In Chapter 6, limitations of this work and future work are discussed. Chapter 7 concludes the thesis.

## 2. Related Work

### Stereotypes and bias

According to the Stanford Encyclopedia of Philosophy [40], implicit bias is

> a term of art referring to relatively unconscious and relatively automatic features of prejudiced judgment and social behavior. … the most striking and well-known research has focused on implicit attitudes toward members of socially stigmatized groups, such as African-Americans, women, and the LGBTQ community.

The occupational gender bias is a conscious or unconscious favoritism towards one gender over another [47] with respect to professions and occupations. Thus some professions can be considered as masculine, neutral, or feminine [48]. For example, nursing is usually stereotyped to be a “feminine” occupation [54].

Biases and corresponding stereotypes are established by past experiences from interactions with people in our society (e.g., by observing friends and parents) [41, 42], influence of cultures [43], and media exposure [7]. Arendt et al. [7] proved that long-term exposure to biased media creates negative implicit attitudes towards the object of bias.

### Gender bias in Wikipedia

Due to the increasing importance of online media, much research [14, 15] is concerned with an assessment of bias on the World Wide Web. A lot of attention was drawn towards biases in Wikipedia and especially gender bias. Reagle and Rhue [12] compared biographical articles from the English Wikipedia edition and the online Encyclopedia Britannica with respect to coverage, gender representation, and article length. Authors concluded that Wikipedia provided better coverage and longer articles. While Wikipedia has more articles on women than Britannica in absolute terms, Wikipedia articles on women are missing more often than are articles on men, when compared to Britannica. Wagner et al. [4] studied coverage of famous women in Wikipedia articles and the way women are portrayed in the online encyclopedia. The authors found that, despite good coverage of famous women in many Wikipedia language editions, the ways in which women and men are portrayed differ significantly. For example, romantic relationships and family-related issues are much more frequently discussed about women than men. Graells-Garrido et al. [16] researched differences between descriptions of men and women biographies in terms of meta-data, network structure, and language. Researchers revealed that articles about men are disproportionately more central than articles about women and the words most associated with men are about sports, while the words most associated with women are about arts, gender, and family. Further research [11] was extended to notability, topical focus, linguistic bias, structural properties, and meta-data presentation which systematize and modify approaches used in [4, 16]. Thus, gender bias on Wikipedia can be assessed using articles with biographies. However, neither of these approaches and measurements can reveal gender bias related to professions.

Much research focuses on the gender gap in women representation among Wikipedia editors. One group of studies focused on measuring the gender gap, other group researched reasons, and consequences.

According to the Wikimedia Foundation survey of editors [50], only 9% of contributors are women. In particular, they encountered from 3% to 20% women in top 10 countries of editors' residence. Hill and Shaw [5] modified the technique used for the estimation of female proportion. While researchers reported slightly higher proportion, i.e., 16.1% female editors on Wikipedia, a huge gender gap remains.

Other research [6, 17, 18, 34] analyzed reasons of the gender gap. Lam et al [34] studied how conflict-related behaviors (e.g., blocks and reverts) affect male and female editors in order to understand why an imbalance might exist. For example, authors found that female editors are reverted more than males. Collier and Bear [6] studied causals of female contributors to stop contributing. They found strong support for the hypothesis that the gender contribution gap is due to responses to conflicts. Hargittai and Shaw [17] found that significant Internet experiences and skills help explain a critical aspect of the variation in the contribution rates among men and women. Thus, higher levels of Internet skills predict much greater probability of contribution for men than women. However, low-skilled men and low-skilled women are equally high unlikely to contribute to Wikipedia. Hinosaar [18] manifested that gender differences in the frequency of Wikipedia use and beliefs about one's competence explain a large share of the gender gap.

Antin et al. [19] studied differences between men and women's editing activity in terms of the number and size of the revisions they make. They found that among the most active Wikipedians men tended to make more revisions than women. However, they also found that the most active women in the sample tended to make larger revisions than the most active men.

Bernacchi [13] performed a visual exploration of the gender issue on Wikipedia by studying articles "Man" and "Woman" in several language editions of Wikipedia. Researcher compared sizes of articles, numbers of edits per article, changes in TOC structures, network structures of related articles and intersection of used concepts, topics coverage, revisions in terms of vandalism and deleted content.

However, profession-related gender bias on Wikipedia and corresponding aspects (e.g., images and mentioned people) remain uncovered.

**Gender bias in image search results and traditional media**

Kay et al. [3] studied the gender bias in image search results for several professions. They find that people's perceptions of gender ratios in occupations are quite precise and manipulated search results have a small significant effect on people's perception towards manipulated ratio. Thus, combining the results with cultivation theory [20, 21, 22] leads to an assumption that long term exposure to images, which depict people of a particular profession, influence the perception of gender ratios of employed people in the profession. Thus, it may trigger implicit gender stereotypes in case of exposure to biased images. Therefore, this research also sets out to study gender bias in images that are part of Wikipedia articles about professions.

Several studies [37, 51, 52, 53] reveal gender bias and stereotypes in broadcasts, popular films, and interviews of sport competitions. Higgs [37] proposed language model-based approach to quantify differences in questions posed to female vs. male athletes. Authors found that journalists in tennis interviews ask male players questions that are generally more focused on the game when compared with the questions they ask their female counterparts. Smith et al. [51] revealed gender inequality in popular films. For example, authors encountered only 29.8-32.8% female characters in films released between 2007 and 2013. Moreover, female characters were more likely than male characters to be shown as sexualized, domesticated. Analysis of the portrayal of occupations across films and prime-time shows revealed that more men than women were depicted with an occupation [52, 53]. Although, women (across more than 300 speaking characters) hold marginally more professional jobs than their male counterparts, women are noticeably absent in some of the most prestigious occupational posts. According to study [53], prime time shows are more likely than family films or children's shows to depict powerful female leaders working across a variety of industry sectors. However, women showed in prime-time shows are still not on par with men in the number of powerful positions held across industries.

**Methods for measuring implicit gender bias**

According to White and White [48] explicit stereotype measures may underestimate occupational gender stereotyping. Thus, in our study only implicit stereotype measures are discussed.

Even though implicit bias might be hard to notice, it can be measured for groups of people in studies. The most common methods and procedures can be divided in two groups: measures of activation and measures of association.

Measures of association, like Implicit-Association test [23] and Go/no-go association task [24], are based on the measurement of association between choices. For example, one can be asked to associate profession names either with group "male and family" or with group "female and science". A score is computed by combining error rate and speed of persons' reaction to the profession classification task. Similar association tasks can be performed for images categorization. Since these methods measure implicit bias of a group of interviewed people rather than existing implicit bias in the information piece, the methods cannot be applied to our dataset. Hence, we will use features, i.e., images and profession names which were preferred by Wikipedia editors' community to depict professions. Since most German profession names are either male or female gender-inclusive, as opposed to gender-free, we will associate names with a male and female group.

Measures of activation, e.g., sentence completion [25] and unprimed lexical decision tasks [26], are based on the activation of a stereotype at that time. For example, one can be asked to continue sentence beginnings that contain person names with a given variant in a sentence completion task. Hence, one can measure stereotypic explanatory biases toward races or genders such as measuring the biases in response to using stereotypic black and white names. These methods measure implicit stereotypes of interviewed people at the moment of the test performance. In our research method, we will use person names mentioned in the profession articles as proxy for gender disparities. We want to know the gender that prevailed among people mentioned in the articles about professions.

## 3. Datasets and Data Collection

**Dataset of profession names.** The list of professions from "Bundesagentur für Arbeit"[3] was used in order to create a list of corresponding male-female pairs of profession names. Hence, the following list of pairs was created: "Lehrer"-"Lehrerin", "Krankenpfleger"-"Krankenschwester", "Entbindungspfleger"-"Hebamme" etc. Also, we extracted neutral names of professions from the initial professions list, e.g., "PR-Fachkraft", "Fotomodell", "Aufsichtsperson". The automatically created lists were cross-validated by two experts.

The following paragraphs describe preprocessing techniques and parsing rules which were applied in order to create the list of corresponding male-female pairs of profession names and the list of neutral names of professions.

The following preprocessing steps were applied:

1. Names which are not professions or occupations and are rather field or activity names were excluded. Thus, all names which consist of substrings "(weiterf","(grundst", "(Tätigkeitsfeld)", "(Staatse" are excluded. The heuristics were applied, because these words indicate one of the following: field name with additional study (German: "weiterführend"), field name with basic study (German: "grundständig"), name of activity field (German: "Tätigkeitsfeld"), field name with the state examination (German: "Staatsexamen").
2. Punctuation errors were fixed and shortenings/abbreviations (e.g., "med.") were converted into full forms.

The remaining strings with profession names mostly consist of a male and a female profession name divided by the symbol "/" (e.g., "Chiefsteward/-stewardess"). Thus, occupations are divided into female and corresponding male profession name by applying the following rules:

1. If a string has one of the following substrings "er/in", "(er/in)", "(e/in)"; then name "first_part_of_name"+"er" or +"e" is assigned to the male title and profession name "first_part_of_name"+"in" is assigned to the corresponding female title. For example, the pair "Lehrer" & "Lehrerin" was created from the string "Lehrer/in".
2. If a string has one of the following substrings "steurer/steurerin", "steward/stewardess", "lotse/lotsin", "amter/amtin", "gehilfe/gehilfin", "arzt/ärztin", "pfleger/schwester", "beauftragter/beauftragte", "anwalt/anwältin", "atrose/atrosin", "purser/purserette", "iker/ikerin", "stuerer/stuerin", "koch/köchin", "aloge/alogin", "ologe/ologin", "pädagoge/pädagogin" ,"logopäde/logopädin", "mann/frau", "experte/expertin"; then the string is divided into male and female name correspondingly. For example, the male-female pair of profession names “Sportpädagoge” and ”Sportpädagogin” was created from the string “Sportpädagoge/-pädagogin”.

[3] Retrieved on 15.06.15 from http://berufenet.arbeitsagentur.de/berufe/berufe-beschreibungen.html, alternative link https://berufenet.arbeitsagentur.de/berufenet/faces/index?path=null/sucheAZ&let=A, file can be accessed from https://github.com/gesiscss/Wikipedia-Language-Olga-master/blob/master/de/Berufsbezeichnungen.txt

3. If a string has one of the following substrings "stellte/stellter", "beauftragte/beauftragter", "schwester/pfleger", "mutter/vater", then the string is divided into a female and a male name accordingly.

Profession names, which end with "kraft", “person”, or “leute”, were inserted into the list of neutral profession names. The remaining profession names were assigned manually. Neutral names are those names which do not impose on gender (gender-exclusive words). Usually, in German grammar it is either plural names (e.g., “Kaufleute”), or female/male/neutral gender words which do not enclose gender of the person (e.g., “Wissenschaftliche Hilfskraft”), or words which are inherited from other languages (e.g., “Model”).

Thus, two lists were created: a list of male-female pairs of profession names and a list of neutral names of professions. The first list consists of 4274 pairs with male and corresponding female form, the second list has 183 entries of neutral names of professions.

Afterwards, all profession names were matched to Wikipedia articles using small Levenshtein distance and ratio. A more detailed description of the matching method is presented in Section 4 (Research Method). In this manner, 885 Wikipedia articles about professions and occupations were found.

**Dataset of person names.** The dataset of people, which are mentioned in the Wikipedia articles about professions, was collected through two separate methods.

Method 1:

All links from the profession articles (outlinks) which are either in Category “Women” (German: “Frau”) or “Men” (German: “Mann”) were saved. The MediaWiki API[4] was used for the aggregation of links. Thus, names of people with a Wikipedia article were collected. As a result, we obtained lists of women and men for each Wikipedia article. In the sequel, this dataset is called LinkDataset.

Method 2:

First, the full texts of profession articles were stored using the MediaWiki API. Second, from the article texts all Entity Names of the class “Person” were mined. For this purpose, we used the library polyglot[5], which implements the Named Entity recognition method proposed by Al-Rafou et al. [27]. Third, for each found person, the gender was identified according to the first name. In order to make gender identification more accurate, several vocabularies were used:

[4] https://de.wikipedia.org/w/api.php

[5] Project page http://polyglot.readthedocs.org/

- vocabulary of the program “gender”[6] by Jorg Michael [28],
- Database Genderizer[7] [29].

Thus, lists of women and men for each Wikipedia article were gathered. In the sequel, this dataset is called PolyglotDataset.

The datasets obtained by these two methods were merged. The PolyglotDataset consists of 3536 men and 723 women and the LinkDataset consists of 2685 men and 347 women. One can see that more men than women are present in both datasets. By merging these datasets, a combined dataset with 5085 persons (4272 men and 813 women) was obtained.

Comparison of the datasets showed that 62% persons (1949 men and 257 women) were present in both datasets, which is 75% of the LinkDataset and 53% of the PolyglotDataset. Thus, 47% of persons (1587 men and 466 women) from the PolyglotDataset are mentioned in the articles but they are not linked to Wikipedia articles. In other words, 466 out of 813 (57%) women and 1587 out of 4272 (37%) men are mentioned, but they are not linked to Wikipedia articles. One can see from these proportions that the majority of women who were mentioned in the articles do not have hyper-links to Wikipedia articles about them, whereas only 37% of mentioned men have no links.

By merging the datasets, we can also estimate the error rate (2.77%) of the gender determination of those persons from the PolyglotDataset who are also in the LinkDataset. This was possible since persons from the LinkDataset are readily gendered.

**Dataset of images.** The MediaWiki API was used in order to retrieve all images from articles. Only images which are wider than 100 pixels were stored, assuming that small images are either icons or too small to recognize the gender. Files of “svg”, “ogg” and “ogv” formats were excluded, since “svg”-files are vector images used for schemas and icons, and “ogg”, “ogv” are video and multimedia text formats. Thus, 906 images from 345 profession articles were collected. The remaining articles do not have (suitable) images.

**Dataset of the German labor market statistics.** Gender-specific employment statistics (German: “Beschäftigungsstatistik”)[8] were obtained from “Statistics of the Federal Employment Agency” (German: “Statistik der Bundesagentur für Arbeit”). The statistics consist of absolute numbers of men and women involved in the profession subgroups as of 30th June 2015. Some examples of profession

---

[6] Project page https://autohotkey.com/board/topic/20260-gender-verification-by-forename-cmd-line-tool-db/ . Python implementation https://pypi.python.org/pypi/SexMachine/ , Dataset in c’t Magazine http://www.heise.de/ct/ftp/07/17/182/

[7] Project page https://genderize.io/

[8] Retrieved from https://statistik.arbeitsagentur.de/Statistikdaten/Detail/201506/iiia6/beschaeftigung-sozbe-bo-heft/bo-heft-d-0-201506-xlsx.xlsx on 19.01.2016

subgroups are these: “8445; (Fremd-)Sprachenlehrer/innen” and “8442; Berufe in der Religionspädagogik”.

Each profession from our dataset was assigned to the profession group according to an accompanying profession classifier[9] (German: “Klassifikation der Berufe 2010 - alphabetisches Verzeichnis der Berufsbenennungen”), i.e., by using the profession encoding number (e.g., “8445x” is encoding of all professions in subgroup “(Fremd-)Sprachenlehrer/innen”, see the previous example). Then, for each profession, the percentage of women in the profession was estimated.

Statistics were obtained for almost all professions with a Wikipedia article, namely for 871 professions. We were not able to obtain statistics for some ambiguous professions/occupations, e.g., “Fachleiter”, “Fachkraft”, “Helfer”.

**Dataset of Google hits.** The numbers of Google search results (hits) were collected for male and female names of professions using the Google Web Search API[10]. Results were retrieved from the API because numbers of Google hits from www.google.de are not reliable. That is, the number of hits can vary, since it depends on custom settings such as used browser, location of user, and Google profile of user.

The search scope was restricted to German language using the corresponding parameter in the API query, as some search terms could also be valid English words, for example.

[9] Retrieved from http://statistik.arbeitsagentur.de/Statischer-Content/Grundlagen/Klassifikation-der-Berufe/KldB2010/Systematik-Verzeichnisse/Generische-Publikationen/Systematisches-Verzeichnis-Berufsbenennung.xls on 19.01.2016

[10] http://ajax.googleapis.com/ajax/services/search/web?v=1.0&hl=de&btnG=Google+Search&q=%22Pflegediensthelfer%22 (depricated), project immigrated to Google Custom Search API https://developers.google.com/custom-search/

## 4. Research Method

First, the dataset of profession names was automatically matched to Wikipedia articles using full profession names from the datasets. Thus, we found articles with the same titles as profession names in the datasets. For each matched article, information regarding its redirection was stored, i.e., whether the original page automatically redirected to another article or not.

Then, we validated if the articles are about profession. For this purpose, all categories an article belongs to were checked. If the article belongs to one of the following Wikipedia categories [11] “Profession” (German: “Beruf”), “(Public) position” (German: “Amt”), “Person by occupation” (German: “Person nach Tätigkeit”), or their subcategories down to 5$^{th}$ depth level, we assume that the article is about profession. Consequently, we restricted all matched articles only to those which are about professions (or occupations).

In order to find more matches, the following semi-automatic method was applied. All titles of articles that were gathered from Wikipedia categories “Profession”, “(Public) position”, “Person by occupation”, and their subcategories down to 5$^{th}$ depth level, were stored. Then the Levenshtein distance and ratio were calculated between each profession name and each Wikipedia article title, gathered in the previous step. The Levenshtein distance measures the minimum number of single-character edits required to change one word into the other [46]. The Levenshtein ratio between words *a* and *b* is defined as follows:

$$LevRatio(a, b) = 1 - \frac{\mathrm{LevDistance(a,b)}}{max\{len(a), len(b)\}}, \quad (1)$$

where $\mathrm{LevDistance(a, b)}$ is the Levenshtein distance between words *a* and *b*, $len(a)$ and $len(b)$ define the length of words *a* and *b*. Thus, the Levenshtein ratio reflects the match proportion of two words. If the Levenshtein distance was at most 2 or the Levenshtein ratio was at least 0.8, the corresponding pairs of words were stored. Hence, we were able to find profession articles with titles written in a slightly different manner than in the original dataset of professions. In order to avoid inappropriate matches (e.g., "Richter" and "Gichter"), all matched pairs were manually checked. Then the new name (article title) was manually assigned to male, female or neutral group of profession names.

Since the matched articles were taken from profession (and occupation) categories, there is no need to further validate them to be about professions.

Thus, a high quality list of Wikipedia articles (885 entries) about professions was collected.

---

[11] https://de.wikipedia.org/wiki/Wikipedia:Kategorien

## 4.1. Redirection Analysis

The first dataset of profession names consists of male-female pairs of profession names. The study reveals how many and which professions are represented by both Wikipedia articles (with male and female job titles) and how many professions are represented at least by one.

If a profession has two articles (with male and corresponding female title) on Wikipedia, we assume that such profession is not biased against men or women, i.e., there is no bias. Therefore, that profession will be associated with a “neutral” group. If a profession has only Wikipedia article with male title, the profession will be associated with a “male bias” group. If it is the other way around, i.e., a profession has only Wikipedia article with female title, the profession will be associated with a “female bias” group.

If redirection takes place then the profession will be associated with the “male bias” group in case of redirection from female to male job title and with the “female bias” group in case of redirection from male to female job title.

The third case is when a profession name redirects to article with neutral name of the profession. We will assign such professions to the “neutral” group. Professions which are only represented via articles with neutral profession names were also associated with the “neutral” group.

Summary of the classification rules are as follows:

- Profession articles with male and female titles do not exist on Wikipedia → no evidence;
- Profession articles with both titles exist (i.e., male and female titles) → **neutral;**
- A profession article with only a male title exists → **male bias;**
- A profession article with only a female title exists → **female bias;**
- A profession article exists, but it automatically redirects to the other gender → bias for redirection target **(**i.e., **female** or **male bias**);
- A profession article with a male title automatically redirects to the neutral form or field name → **neutral;**
- A profession article with a female title automatically redirects to the neutral form or field name → **neutral;**
- A profession article with a neutral title exists → **neutral.**

After assigning professions to the redirection bias groups, one can assess their size. In other words, one can find whether professions are more likely to be presented via male, female or neutral profession names. For instance, if one observes more professions in the male bias group, it means that there are more professions which are represented only via articles with a male title in Wikipedia.

Next we check if predominance of one bias group can be explained by popularity of profession names on the World Wide Web or by German labor market statistics. That is, Wikipedia could reflect and inherit gender bias from other media or gender inequality of German labor market.

#### 4.1.1. Google Results

Initially, we investigate whether phenomena observed on Wikipedia are specific to Wikipedia or if they can be explained by popularity of profession names on the Web. Thus, we hypothesize that profession titles that appear on Wikipedia are more popular on the Web, meaning that there are more sources on the Web about them than about the corresponding profession titles of the opposite gender. In other words, one would observe more search results for those profession titles.

Therefore, we look into the number of results (hits) returned by the Google Search Engine for each profession title. To this end, the Google Web Search API was used, which, besides ranked results also returns the number of Google search results for a specific query. For each profession, the number of results (hits) was stored for female and male job titles.

In order to assess if the difference between Google hits for female and male titles is significant, the two-sided Wilcoxon-Mann-Whitney rank-sum test was performed between hits for female and male job titles.

Next, we checked whether one can describe the relationship between the number of Google hits and the redirection bias of a profession. First, for each profession, the normalized difference was estimated using the following formula:

$$Normalized_difference_i = \frac{Hits_{male_i} - Hits_{female_i}}{Hits_{male_i} + Hits_{female_i}}, \quad (2)$$

where $Hits_{male_i}$ is the number of results (hits) returned for male title of profession $i$, $Hits_{female_i}$ is the number of results (hits) returned for the corresponding female title of profession *i*.

Then, we checked the possibility of predicting the redirection bias using Google hits for profession names. As described earlier, the professions were classified into three redirection bias groups: male, female, and neutral. In the next step, we fit models which predict the redirection bias group of profession.

Two logistic regression models were fitted. The first logistic regression model uses as a dependent variable the state of being in the male bias group; the second model deals with female bias instead. In other words, the models will predict whether a profession has male or female bias. The regression functions for both models are given by:

$$p_i = \frac{1}{1+e^{-(\beta_0+\beta_1 x_{1,i}+\beta_2 x_{2,i})}}, \quad (2)$$

where $x_{1,i}$ is the normalized difference of Google hits of profession $i$, $x_{2,i}$ is the Google hits for male title of profession $i$. In other words, independent variables in both models are the following: the normalized difference of Google hits; Google hits for the male job title.

### 4.1.2. German Labor Market

This study examines whether phenomena observed on Wikipedia can be explained by labor market statistics. One could hypothesize that professions with more women might only have a female article, whereas professions with more men might only have a male article.

For our analysis, the German labor market statistics were used. The statistics consist of numbers of people involved in profession subcategories according to a classification of professions (German: "Klassifikation der Berufe 2010"). Each profession was associated with an appropriate profession subcategory according to the aforementioned classification. Thus, each profession was coupled with the corresponding number of employed people per gender. Since some of the professions are too ambiguous (e.g., occupation "Leiter"), there is no labor market statistics for them. Statistics from the German labor market statistics were associated with 859 out of 873 professions existing on Wikipedia.

For each profession, the percentage of women involved was obtained. Then, the dependence between the percentage of women involved in the profession and the redirection bias group of profession was checked. The null hypothesis is that two sets of measurements are drawn from the same distribution, i.e., the percentage of women involved in profession of the first bias group is drawn from the same distribution as the percentage of women involved in the profession of the second bias group. The alternative hypothesis is that values in one sample are more likely to be larger than the values in the other sample.

The Wilcoxon rank-sum test was used for each pair of the bias groups (male-female bias, male-neutral bias, neutral-female bias) in order to find profession groups which show significant differences in the percentage of employed women. The Bonferroni correction [30] was used in order to control the family-wise error rate.

To describe the relations between the bias groups using the percentage of women involved in professions, logistic regression models were fitted. Analogously to the analysis of Google results, we predict whether a profession is in the female\neutral\male bias group. The explanatory variable is the percentage of women involved in a profession. Since not all bias groups show significant difference between distributions of values, we will fit our model only between those which do.

### 4.2. Images Analysis

All images from the profession articles were extracted using the MediaWiki API. Then, the following files were excluded:

a) images of .svg format, because they are vector image that are used for Wikipedia logos or for drawing some schemas;
b) files of .ogg and .ogv formats, because they represents video, audio and text media files;
c) images with width less than 100 pixels.

In this manner, we extracted 906 images.

In order to identify gender of people on the photo, a CrowdFlower task was set up. Interfaces designed for the CrowdFlower task are presented in **Figure 2**. Depending on the number of depicted people, the CrowdFlower workers received slightly different questions.

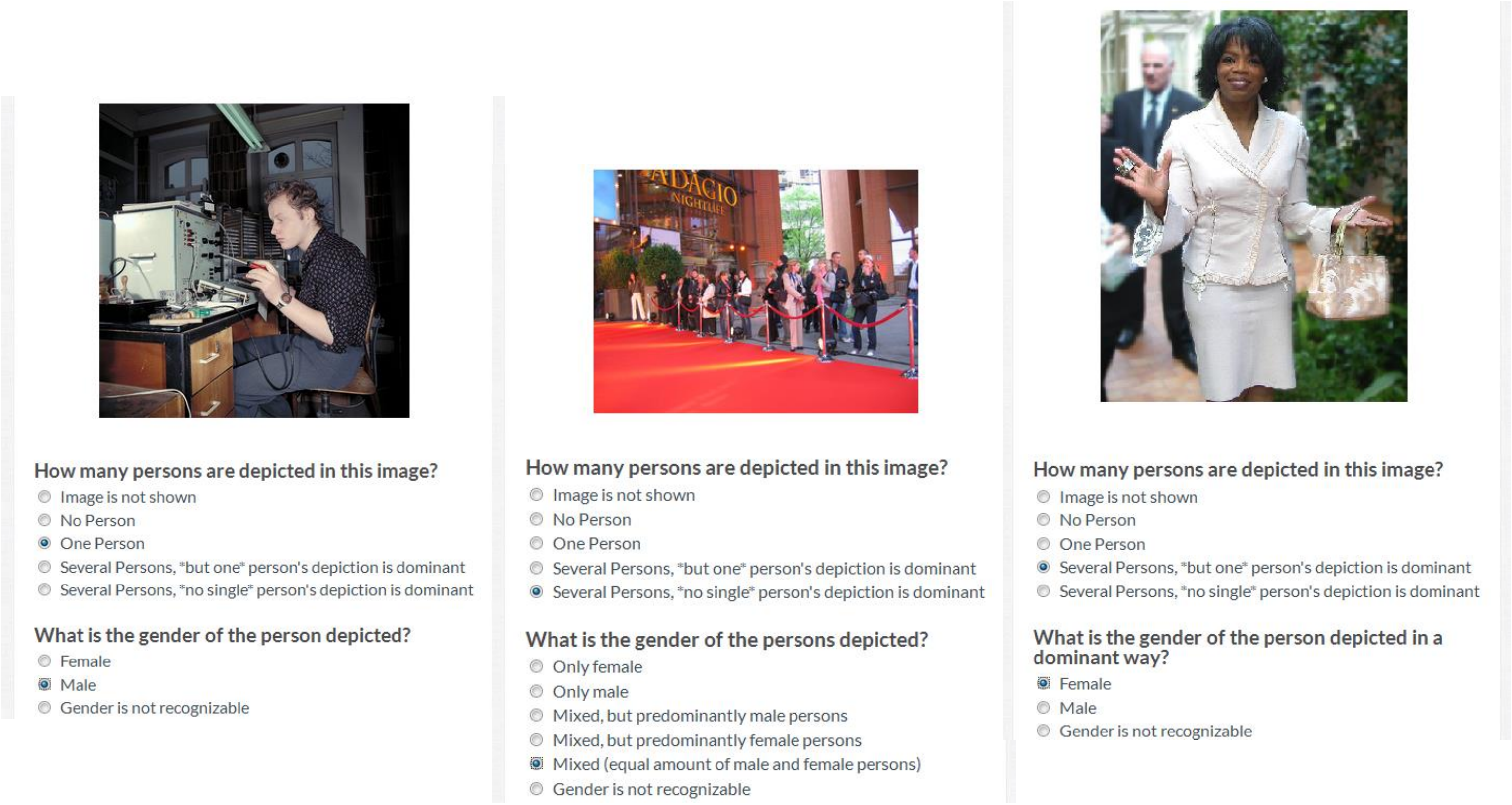

Figure 2a Figure 2b Figure 2c

**Figure 2:** Examples of questions on CrowdFlower platform; figure **a)** shows a question for the case when there is only one depicted person; figure **b)** shows a question if there is more than one person depicted and no single person's depiction is dominant; figure **c)** shows a question for the case when there is more than one person depicted and one single person's depiction is dominant.

First, CrowdFlower workers should identify whether the photo shows people or not, by choosing one of the following answers: “No Persons”, “One Person”, “Several Persons, *no single* person’s depiction is dominant”, “Several Persons, *but one* person’s depiction is dominant”. Thus, if the image depicted more than one person, workers were asked to identify whether one person was the main object of an image (i.e., the person is depicted in a dominant way). We add such constraint since images can depict several people and we are interested only in the gender of the main person. Figure 3 represents several examples of Wikipedia images which depict more than one person, but one person’s depiction is dominant.

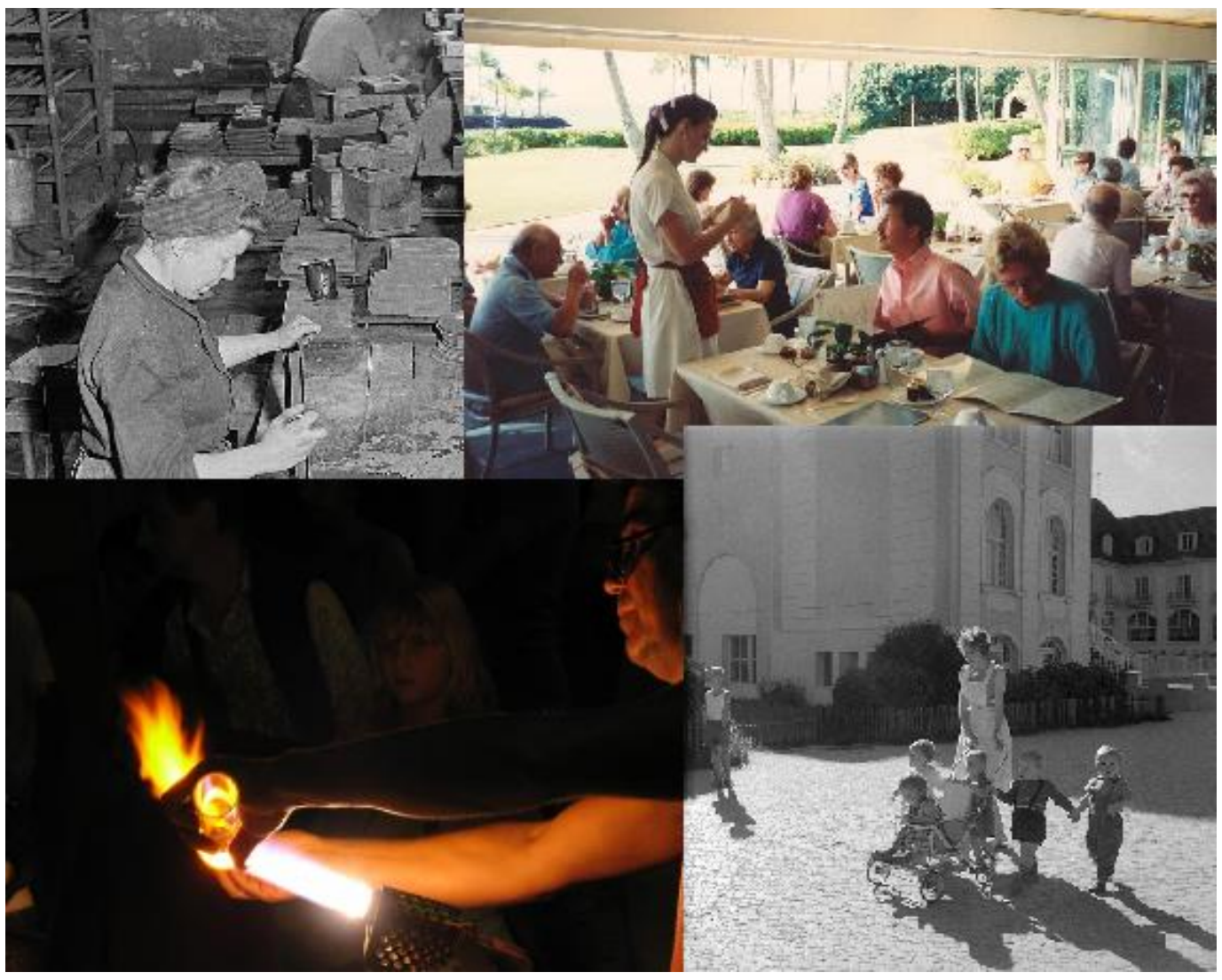

*Images from left upper corner:*
Kernmacherin bei der Arbeit © Deutsche Fotothek / CC-BY-SA-3.0 de / http://creativecommons.org/licenses/by-sa/3.0/de/deed.en
A waitress taking a breakfast order at Kahala Hilton Hotel © Alan Light / CC-BY-2.0 / http://creativecommons.org/licenses/by/2.0
Glasbläser am Zenitbrenner
Erzieherin mit Kindern (1963) © Deutsche Fotothek / CC-BY-SA-3.0 de / http://creativecommons.org/licenses/by-sa/3.0/de/deed.en

**Figure 3:** Wikipedia images from the profession articles; images with several persons, where one person’s depiction is dominant. Image names (from left upper corner): “Kernmacherin bei der Arbeit“, „Waitress taking an order“, „Glasbläser“, „Erzieherin mit Kindern (1963)“.

Second, for every image with exactly one person, workers were asked to identify the gender (Figure 2a) of that person. For every image with more than one person where one is dominant, workers had to identify only the gender for that dominant person (Figure 2c). Otherwise, they were asked to identify the gender of the majority of people on the photo (Figure 2b). If the gender is not recognizable or if the photo shows about the equal number of men and women, they should choose the corresponding variants (Figure 2b).

Each image was classified by at least three different workers. In order to control the reliability of all responses, the accuracy threshold for workers was defined as 70% in the setting of the CrowdFlower task. We also checked how consistent the responses of the workers were for the whole project; Fleiss' kappa score was utilized.

We manually labeled 15% of images such that one of 10 images shown to a CrowdFlower contributor would be from the labeled set. Since we know the “right answer” for the labeled images, the accuracy of contributors is assessed based on their answers for those images. In other words, accuracy score is the percentage of correct answers for the labeled images. When a contributor starts our CrowdFlower job his or her accuracy score equals 100%. The accuracy score is recalculated after receiving answers for 10 images by CrowdFlower system. If the answer of the contributor does not match “the right answer”, contributor’s job accuracy goes down. If the accuracy of worker falls below the accuracy threshold, such contributor will be removed from the job and his or her answers will not be taken into account. Moreover, in order to start CrowdFlower task, workers should pass the test tasks, where all 10 images are chosen from the labeled set.

For the whole CrowdFlower task the reliability of agreement Fleiss' kappa [33] was estimated using the following formula:

$$\boldsymbol{\kappa} = \frac{\bar{P} - \overline{P_e}}{1 - \overline{P_e}}\,, \qquad (3)$$

where $\bar{P}$ refers to the mean of $P_i's$ and $P_i$ refers to the extent to which workers agree for the *i*-th image. In other words, $P_i$ defines how many worker--worker pairs are in agreement, relative to the number of all possible worker -- worker pairs and is calculated using the following formula:

$$P_i = \frac{1}{n(n-1)}\left[\left(\sum_{j=1}^{k} n_{ij}^2\right) - n\right]\,, \qquad (4)$$

where $n_{ij}$ refers to the number of raters who assigned the *i*-th image to the *j*-th category, *n* refers to the number of answers per image (i.e., three in our case), *k* refers to the number of categories.

$$\overline{P_e} = \sum_{j=1}^{k} p_j^2, \qquad (5)$$

where $p_j$ refers to the proportion of all assignments which were made to the *j*-th category of images.

Fleiss’ kappa allows us to measure the degree to which the observed amount of agreement among workers exceeds what would be expected if all workers made their choices completely randomly. In other words, it can give us a clue on how consistent responses of the workers are.

After gathering all answers for all images, the majority answer was used to label the image.

The images were grouped in a way that: images depicting one male, one dominant male, and images with male majority were assigned to one single group “male”; images depicting one female, one dominant female, and images with female majority were assigned to one single image group “female”; images where gender was not recognizable and with equal number of men and women were assigned to two separate groups “gender is not recognizable” and “mixed, equal amount of male and female”.

In order to check whether images from the profession articles reflect **labor market** statistics, the following analysis of images was performed. Professions were divided into two groups: professions with female majority and professions with male majority according to the labor market statistics. Thus, professions with more than 50% women were in one group and professions with more than 50% men were in another group. Then, we tested the statistical significance of the difference between distributions of image groups using chi-square independence tests with Monte Carlo p-value simulations [31]. The Monte Carlo p-value simulations were used since some image categories can have small numbers, such that p-value can be unreliable. The null hypothesis is: category of image and gender of majority in profession are independent. The alternative hypothesis is: category of image and gender of majority in profession are not independent.

We also checked how distribution of image categories would look like if one restricts professions to those with more than 70% men or women, respectively. Analogously, chi square test was used in order to check whether the difference is significant.

Next, the strength of relation was checked between the number of images depicting a particular gender in the article and the labor market statistics of profession. The Spearman’s rank correlation (with correction for ties [39]) was utilized.

We also checked whether distributions of image categories are significantly different for **articles with male, female and neutral titles**. The images were grouped according to the gender of article titles. Then chi-square independence tests with Monte Carlo p-value simulations were performed. Thus, we test the statistical significance of the difference between distributions of image categories of these article groups. The null hypothesis is: category of image and gender of article title are independent. The alternative hypothesis is: category of image and gender of article title are not independent.

Analogously, we checked whether distributions of image categories are significantly different for professions which were assigned to different **redirection bias groups**.

If the test showed a statistically significant difference in the image composition between different article/profession groups, the groups which show the significant difference were revealed. For this purpose pair comparisons (i.e., post-hoc tests) of all groups were performed. To control for the family-wise error rate, the two stage p-value correction of Benjamini-Hochberg [32] was utilized.

## 4.3. Mentioned people analysis

The dataset of person names were used for this analysis. Articles that did not mention any persons were excluded from the analysis. Then, for each article, the ratio of mentioned men was estimated.

First, the dependency between the ratio of mentioned men and the gender of the article title was checked. Thus, we examine if:

a) articles with **male title** have higher ratio of mentioned men than articles with **neutral title**,
b) articles with **male title** have higher ratio of mentioned men than articles with **female title**,
c) articles with **neutral title** have higher ratio of mentioned men than articles with **female title**.

To this end, three Wilcoxon rank-sum tests (i.e., one for each pair of article groups) were performed. The following data was compared: ratios of mentioned men in articles with female and male title; ratios of mentioned men in articles with female and neutral title; ratios of mentioned men in articles with male and neutral title. The null hypotheses are that two sets of ratios of mentioned men are drawn from the same distribution. The alternative hypotheses are that values in one set are more likely to be larger than the values in the other set. The two stage p-value correction of Benjamini-Hochberg was used in order to control for the family-wise error rate.

Second, we checked the dependency between the ratio of mentioned men and **redirection bias group** of profession. Thus, analogously to the previous analysis, the professions were grouped and then Wilcoxon rank-sum tests were applied between ratios of mentioned men in each of the groups. According to the redirection analysis we have three redirection bias groups (male, female and neutral), thus it requires three tests and p-value correction. Analogously, the two stage p-value correction of Benjamini-Hochberg was applied for the family-wise error rate control.

Third, we examined whether the ratios of mentioned men in the profession articles reflect **labor market** statistics. The professions were divided in two groups: professions with more than 50% men and professions with more than 50% women in labor market. If ratios of mentioned men in the profession articles reflect labor market, one would observe high ratio of mentioned men in the group of professions with male majority and low ratio of mentioned men in the group of professions with female majority. We checked whether the difference is significant between the groups of professions using Wilcoxon rank-sum test.

Next we analyze the strength of the relation between the percentage/number of mentioned men/women in an article and the percentage/number of employed men/women in a profession. Thus the Spearman's rank correlation (with tie correction) was calculated between:

- the percentage of mentioned women in profession article and the percentage of employed women in the profession;

- the number of mentioned men in article and the number of employed men in the profession;
- the number of mentioned women in article and the number of employed women in the profession;
- the number of mentioned men in article and the number of employed women in the profession;
- the number of mentioned women in article and the number of employed men in the profession;
- the number of mentioned men in article and the number of employed people in the profession;
- the number of mentioned women in article and the number of employed people in the profession.

Wikipedia articles usually reflect not only on current state of profession, but also the historical past of it and famous people involved in different points in time. Therefore, the mentioned people in articles were limited to those who reflect current labor force. The pool of people was restricted to those who were born after 1960. In order to get dates of birth of mentioned people, the SPARQL queries on DBpedia were applied. The predicate “birthDate” values of persons, who were mentioned in articles, were queried. We were able to find DBpedia objects according to their article id in Wikipedia, i.e, the scope of mentioned people was limited to those with a Wikipedia article. For persons who do not have DBpedia entry, the first line of their Wikipedia articles was parsed in order to retrieve birth date.

# 5. Results

## 5.1. Redirection Analysis

In the set of 4457 (4274 male/female and 183 neutral) professions, 991 Wikipedia articles and 820 redirects (Table 1a) were encountered. Further analysis of the Wikipedia articles revealed that only 869 articles were about professions, among them are the following: 815 profession articles with male title, 18 profession articles with female title and 36 articles with neutral name of profession (Table 1b). When corresponding male to female profession names are combined, one will encounter eight professions (i.e., 16 articles) that have corresponding articles with female and male names of professions. These eight professions according to our classification rules (see Subsection 4.1.Redirection analysis) were assigned to the neutral bias (redirection) group. Articles with neutral name of profession (36 cases) were also assigned to the neutral bias group.

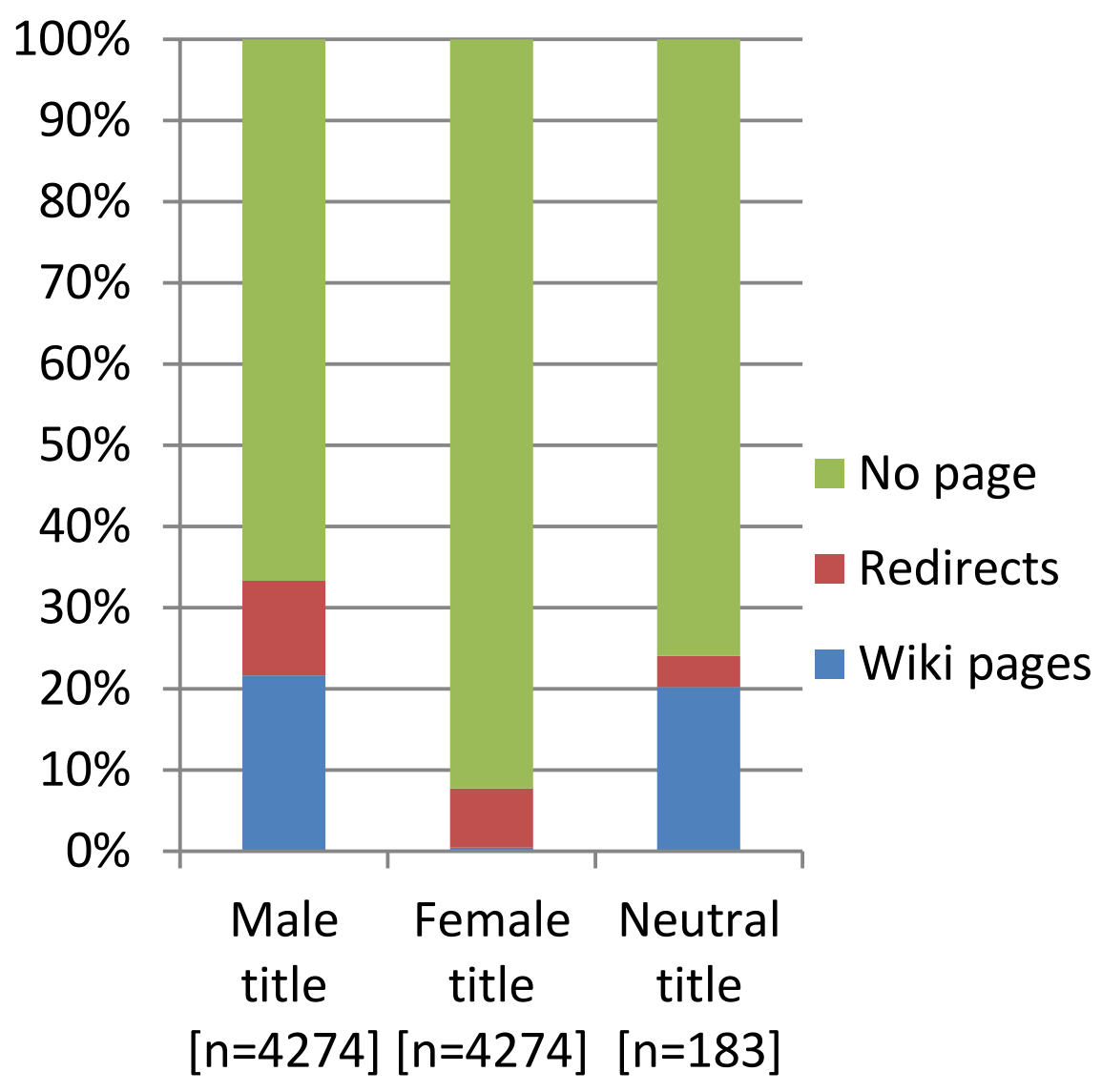

**Figure 4:** The percentage of profession names with male/female/neutral job titles that had Wikipedia page, had no page or redirected to another page. Blue color represents proportion of articles about professions; red color represents proportion of encountered redirections from a seed page; green color represents profession names which have neither page nor redirect.

One can see from Figure 4 that most pages about professions on Wikipedia have male titles and there are only few articles with female titles. There are also few articles about profession with neutral profession names. Hence, at first glance, Wikipedia community is more male profession-oriented.

Among 820 redirects (Table 1c) the following were found: 503 redirects from male label of profession, 310 redirects from female label of profession and seven redirects from neutral label of profession. For example, if one would go to the Wikipedia article “Lehrerin”, one will be automatically redirected to the article “Lehrer”, and thus, one never reaches a Wikipedia page “Lehrerin”. Female professions that redirect to male professions or have only male profession articles were assigned to male bias group. Thus, we identified 807 professions that have only male profession article and five more were found from the set of articles with female title that redirect to articles with corresponding male title. Male professions that redirect to female professions or have only a female profession article are identified as female bias group (six cases). Professions were also assigned to the neutral bias group if the following conditions were encountered simultaneously: profession has only one article (either with male or female title) and corresponding job title of opposite gender redirects but not to articles with male or female job title (11 cases).

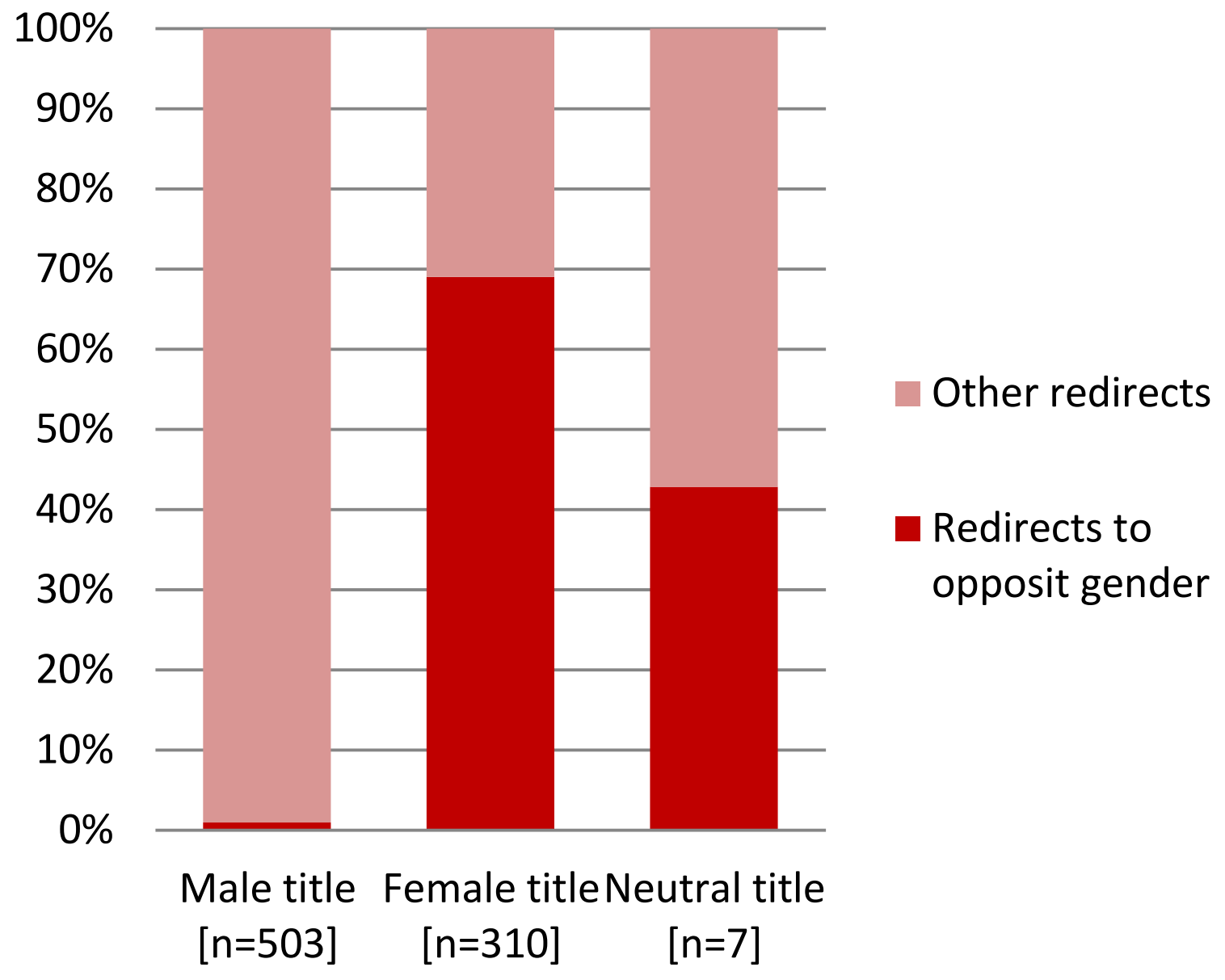

**Figure 5:** The percentage of redirections from male/female/neutral job titles to articles with corresponding job titles of opposite gender and to other articles. Dark red color represents proportion of redirects from seed page to opposite gender, e.g., middle bar shows redirections from pages with female titles to articles with male titles. Dark red color in “neutral title” bar represents redirects to articles with male titles; we encountered 0 redirects to articles with female titles.

One can see from Figure 5 and Table 1c a strong imbalance in the number of redirects between groups of female, male and neutral profession names. Redirects from female to male profession articles are more common than from male to female. Pages with neutral profession names do not redirect to articles with female title and there are three cases with redirects to articles with male title.

Combination of redirects and existing pages of corresponding male-female profession names revealed 812 male bias cases, 6 cases of female bias, and 55 neutral bias cases. Meaning that 812 professions have only an article with male title or a Wikipedia page with female title redirects to an article with the corresponding male title, 6 professions have only an article with female title or a Wikipedia page with male title redirects to an article with corresponding female title, 55 professions have either an article with neutral title or articles with both male and female titles. Thus, we observe evidence for gender disparity among article titles chosen by Wikipedia community editors.

Next, we want to know whether preference of male titles over female titles is a specific phenomenon of Wikipedia. Therefore, we will check how popular male and female profession titles are on the Web. Then we will check whether the preference is aligned with the popularity of professions among men or women on the labor market.

| | All wiki pages | Validated pages about profession | Other pages |
|---|---|---|---|
| Masculine | 934 | 815 | 119 |
| Feminine | 20 | 18 | 2 |
| Neutral | 37 | 36 | 1 |

Table 1b

| | All | Wiki pages | Redirects | No page |
|---|---|---|---|---|
| Masculine | 4274 | 934 | 503 | 2837 |
| Feminine | 4274 | 20 | 310 | 3944 |
| Neutral | 183 | 37 | 7 | 139 |

Table 1a

| | All redirects | Redirects to opposite gender | Other redirects |
|---|---|---|---|
| Masculine | 503 | 5 | 498 |
| Feminine | 310 | 214 | 96 |
| Neutral | 7 | to M:3, to F:0 | 4 |

Table 1c

**Table 1a** represents the number of found articles and encountered redirections on Wikipedia from the set of German professions. Column “Redirects” corresponds to existing Wiki articles which one would reach via the automatic redirection from the initially requested page. Column “Wiki pages” corresponds to existing articles which one would reach without redirection, thus we can reach articles directly. **Table 1b** represents the number of Wikipedia articles which were identified as pages about profession (see Section 3.

Datasets and Data Collection). **Table 1c** represents the number of different redirections. Column “Redirects to opposite gender” stands for cases where one would be automatically redirected to articles of opposite gender when requesting Wikipedia page with profession name of selected gender. Column “Other redirects” stands for redirects to fields of professions, or articles with titles which are synonyms, or articles with neutral name of profession. For example, Wikipedia article “Dressman” redirects to neutral name of profession, i.e., “Model”.

One can see that most pages about professions on Wikipedia have male title and there are only few articles with female and neutral titles. Moreover, there are much more redirects from female to male profession articles than from male to female.

### 5.1.1. Google Results

To examine whether the phenomenon observed on Wikipedia is specific to Wikipedia or it can be explained by the popularity of profession names on the Web, we looked into numbers of sources which exist for male and female profession titles correspondingly.

Our hypothesis is that profession titles with a Wikipedia article are more popular on the Web than corresponding profession titles of opposite gender without a Wikipedia article. In other words, one would observe more sources for male profession titles than corresponding female titles for those professions with only a male article on Wikipedia (i.e., male bias group). One would also observe more sources for female than male titles for professions with only a female article on Wikipedia (i.e., female bias group). This would mean that choices of Wikipedia editors reflect the popularity of profession names on the Web.

First, we compared the number of Google results (hits) for male and corresponding female names of professions. Figure 6 shows the distribution of Google hits for profession names. One can see that male profession names tend to be more popular (in terms of numbers of results) on Google than female names. Most female profession names have fewer Google results than the corresponding male profession names, for instance, “Professor” has more Google results than the corresponding female job title “Professorin” (Figure 6). There are only few professions for which the female profession names are more popular than the male ones. For example, “Hebamme” has much more Google results than the corresponding male name “Entbindungspfleger”.

According to the results of Wilcoxon rank-sum test between distributions of Google hits of male and female profession names, the difference is significant ($p \ll 0.0001$, $z = 28.6$) and Google hits of male profession names are more likely to be greater than respective values of female profession names. This indicates that the German speaking web is biased towards male profession titles, i.e., for most of professions one can find more sources for male than female profession names, which could be also reflected on Wikipedia.

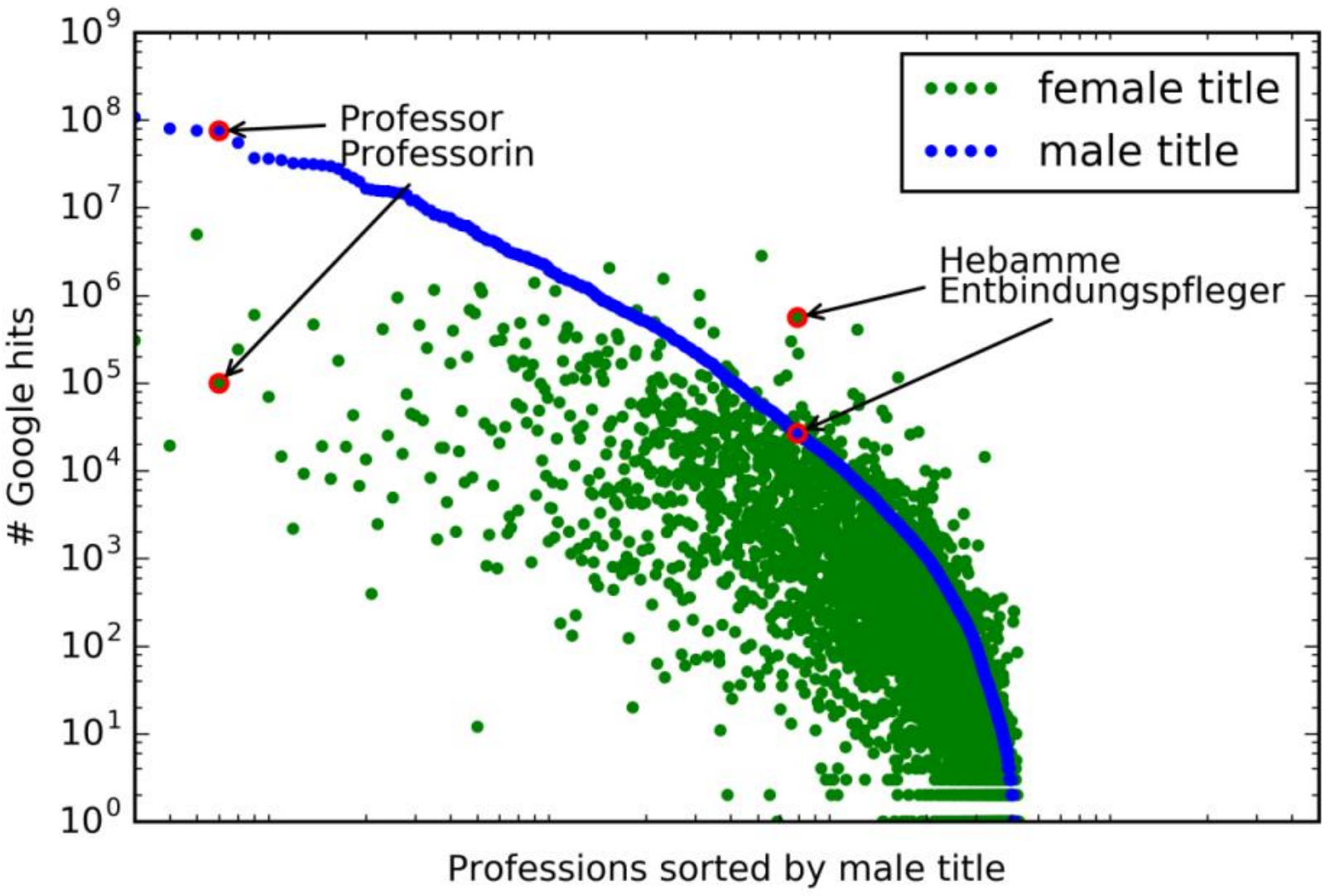

**Figure 6:** Distribution of Google hits for profession names, professions are sorted according to Google hits of male profession name. One can see that most female profession names have fewer Google results than corresponding male profession names.

The next analysis aims at checking the difference between Google hits of professions from male, female, and neutral redirection bias groups. For each profession, the normalized difference between Google hits of male and female name of a profession were estimated using equation 2 (Subsection 4.1.1). Figure 7 shows the distribution of normalized differences between Google hits of male and female profession names that are grouped by redirection biases. One can see that all professions from the female bias group have a negative normalized difference of Google hits. That means that these professions have more Google results for female profession names than male ones. For example, if one queries for “Entbindungspfleger”, Google search engine returns about 100.000 results. At the same time, for corresponding female form “Hebamme”, it returns almost one million results.

At first glance, a female profession should be extremely popular in order to make it to Wikipedia, since median normalized difference of Google hits for female bias group equals -0.86. In other words, ½ of the female bias professions have at least 13 times greater number of Google hits for female professions name than for male name. Next, the differences between male and female bias groups were checked on whether the relationship can be described between the Google hits and the bias group of profession. In other words, we checked the impact of the normalized difference of Google hits (as well as Google hits for male job title) on the bias group of profession. To model this influence, logistic regressions were trained to adapt the bias category of a profession for a one-unit change in the normalized Google difference of the profession and Google hits of male job title. Two logistic regression models were fitted.

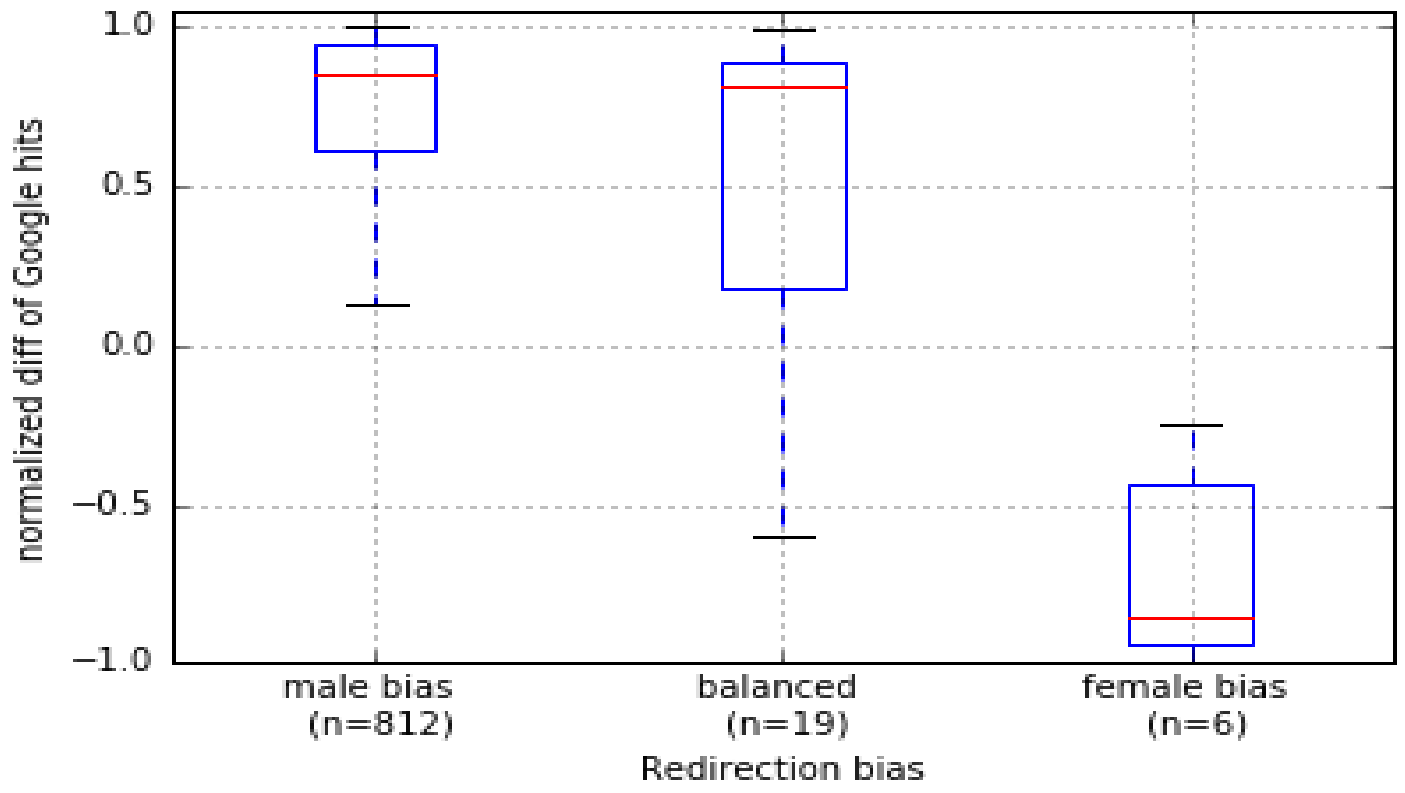

**Figure 7:** Distribution of normalized differences between Google hits of male and female profession names. Professions are grouped according to redirection bias. Male bias stands for cases where only an article with male profession name exists, or a page with female profession name automatically redirects to the respective male article. Female bias stands for cases where only an article with female profession name exists, or a page with male profession name automatically redirects to the respective female article. Balanced stands for cases of a neutral bias, where both male and female articles exist. One can see that all professions from female bias group have a negative normalized difference of Google hits, whereas all professions from male bias group have a positive normalized difference of Google hits.

The first model predicts female bias and the second model predicts male bias. The independent variables in both models are these: i) the normalized difference of Google hits; ii) Google hits for male job title. The following regression parameters (Table 2) were learned:

- The first model with $\beta_0 = 2.41^{***}$, $\beta_1 = 2.44^{***}$ and $\beta_2 = 0.0$ (is not significant);
- The second model with $\beta_0 = -5.55^{**}$, $\beta_1 = -5.93^{**}$ and $\beta_2 = -2.07\ 10^{-5}$ (is not significant).

Results of the logistic regression models fitting reveal that with the help of the normalized difference of Google hits we can predict if a profession has a male or female redirection bias. Logistic regression coefficients reveal that with a one-unit increase in the normalized difference of Google hits, odds of a profession having a male bias increases by a factor of 11.48, whereas the probability of having a female bias decreases. In other words, it is more likely that professions have only a Wikipedia article with female profession title (female bias group) when the female profession name has greater number of Google hits than the male profession name. The results of the logistic regression models fitting are shown in Table 2.

| | Coef. | p | 95% Conf. int. |
|---|---|---|---|
| Model 1: | | Accuracy: 0.971 | Pseudo R-squared: 0.21 |
| Normalized Google difference | 2.44 | 0.0000 | [ 1.68, 3.20 ] |
| Google hits for male name | 0.00 | 0.9949 | [ 0.00, 0.00 ] |
| (Intercept) | 2.41 | 0.0000 | [ 1.92, 2.90] |
| | | | |
| Model 2: | | Accuracy: 0.995 | Pseudo R-squared: 0.62 |
| Normalized Google difference | -5.93 | 0.006 | [ -10.13, -1.73 ] |
| Google hits for male name | -2.075e-05 | 0.558 | [ -9.02e-05, 4.87e-05 ] |
| (Intercept) | -5.5543 | 0.001 | [ -8.872, -2.236] |

**Table 2:** Results of the best fitted logistic regression models. Model 1 stands for logit model with binary outcome of profession being in the female bias group or not. The coefficient for "Normalized Google difference" reveals that, we will see 11.48 factor increase in the odds of being in a female bias group for a one-unit increase in "Normalized Google difference" score, since exp(2.44) = 11.48. Model 2 stands for logit model with binary outcome of profession being in the male bias group. The coefficient for "Normalized Google difference" reveals that, we will see 0.0026 factor increase (basically 1/0.0026=384 factor decrease) in the odds of being in a male bias group for a one-unit increase in "Normalized Google difference" score since exp(-5.93) = 0.002654. For example, both regression models will predict a female bias when the value of normalized Goggle difference equals -1 and male bias when the value equals 1.

Our findings reveal that the German speaking web is biased towards male job titles, meaning that for most of the professions one can find much more sources for male than female job titles. The results of analysis of professions with Wikipedia articles reveal that professions from the female bias group have a significantly lower normalized Google difference than the professions in the male bias group. In other words, Google hits are lower for male than corresponding female job titles in professions that have only an article with a female job title as an article title. At the same time, Google hits are greater for male than corresponding female job titles in professions that have only an article with a male job title as an article title. However, the balanced bias group does not show the significant difference with the male and female bias groups. The normalized Google difference of the balanced group varies from 1 to -0.6 (Figure 7), which means that professions in the group can have popular either female or male job titles.

### 5.1.2. German Labor Market

This analysis aims to study whether gender bias in profession titles on Wikipedia can be explained by German labor market statistics. We hypothesize that professions with more women will have only a female page (i.e., professions from the female bias redirection group), whereas professions with more men will have only a male page (i.e., professions from the male bias redirection group) on Wikipedia.

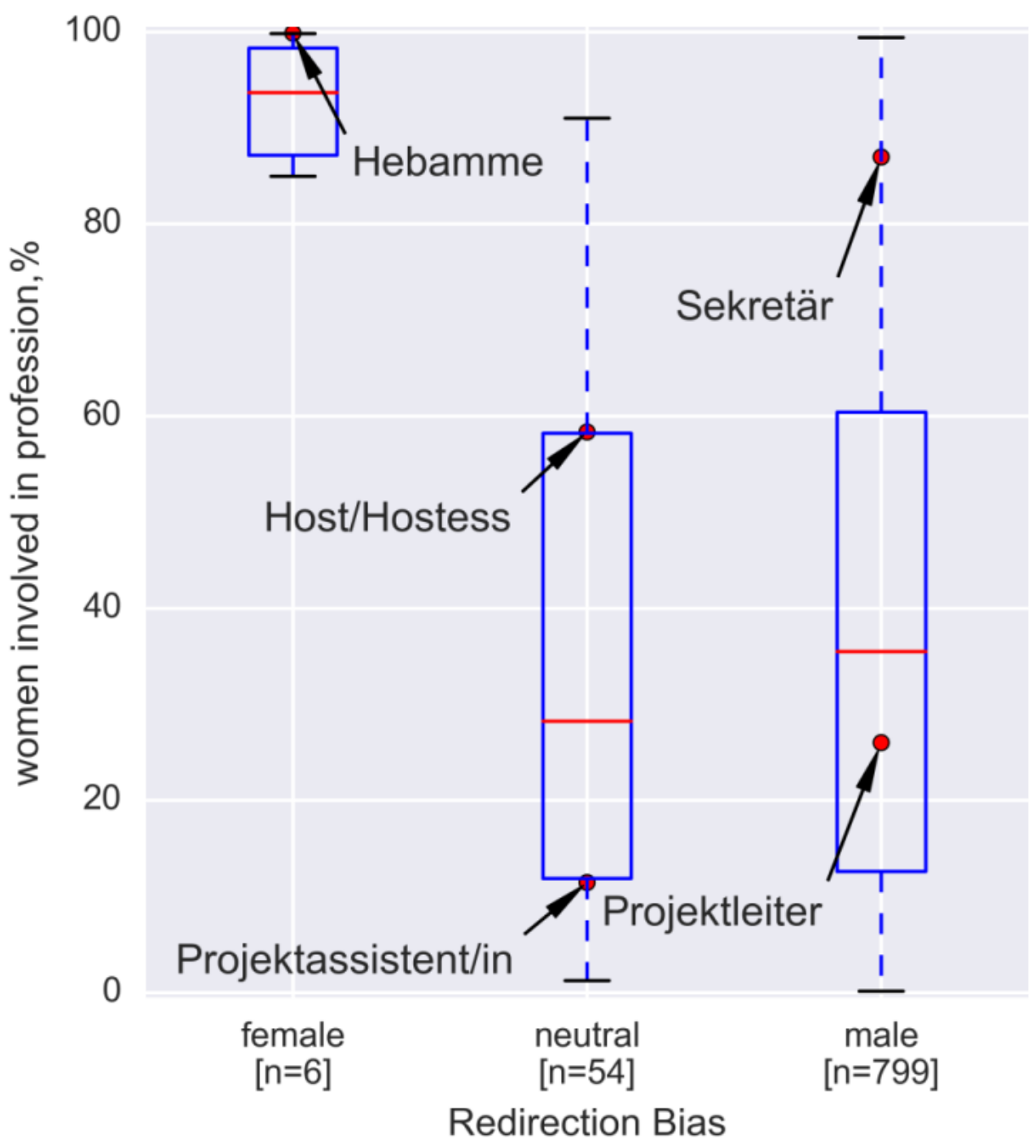

**Figure 8:** Distribution of percentage of women involved in professions (German labor market); data is grouped by redirection bias of profession. The statistics from the German labor market were associated with 859 professions (799 from male redirection bias, 54 from neutral redirection bias and 6 from female redirection bias). Since some of the professions are too ambiguous (e.g., occupation "Leiter"), there is no labor market statistics for them.

One can see that professions with a female redirection bias on Wikipedia are indeed dominated by women in the German labor market, while there is no clear relation with the labor market statistics for the other types of redirection bias.

First, the dependence between the percentage of women involved in a profession and the redirection bias was checked. Figure 8 shows the percentage of women involved in professions; data is grouped by redirection bias. One can see that professions in the female bias group have between 82 and 100 percent of women (median is 93.7) and the male bias professions have from 0 to 100 percent of women, where half of professions in male bias group have between 15 and 60 percent of women. The neutral and male bias groups show similar distributions of women involved in a profession, with slightly different median values, 28.3 and 35.5 percent of women, respectively. This suggests that potentially there is no significant difference between groups of male and neutral bias in terms of employed women.

Next the dependence was tested between the percentage of employed women in the German labor market and the redirection bias. The results of Wilcoxon rank-sum tests suggest a statistically significant difference (the alpha according to the correction method is 0.0167) between the underlying distributions of:

1) the percentage of women involved in male bias professions and the percentage of women involved in female bias professions ($z = -3.32$, $p<0.001$);
2) the percentage of women involved in neutral bias professions and the percentage of women involved in female bias professions ($z = -3.35$, $p<0.001$).

There is no statistically significant difference between the underlying distributions of the percentage of women involved in professions which are in the male bias group and the percentage of women involved in professions which are in the neutral bias group ($z = 0.83$, $p> 0.1$).

Thus, there are statistically significant differences between the female bias group and other groups, in terms of percentage of women involved in a profession. Professions which are represented by only articles with female titles (i.e., female bias) have significantly higher percentage of employed women than other professions, i.e., between 82 and 100 percent of women, whereas professions that are represented by only article with male title tend to have from 10 to 60 percent of women.

Nevertheless, one can see from the Figure 8 that many professions with 80-100 percent of women (in the labor market) are in the male bias group. For example, profession "Gesundheits- und KrankenpflegerIn" encounters about 85% women and the Wikipedia page with female title of profession "Gesundheits- und Krankenpflegerin" automatically redirects to the respective male form "Gesundheits- und Krankenpfleger", hence the profession is in the male bias group. Another example is profession "Sekretär/in" which encounters about 88% women in Germany and has only male article "Sekretär". Thus, we cannot claim that Wikipedia community decided to use only female profession names as article titles for professions with more than 80 percent of women in the labor market. We can only conclude that existing professions with only a female profession article are more likely to have higher percentage of women in the labor market than professions with only a male profession article.

Second, the logistic regression model was fitted in order to describe relations between the redirection bias groups using the percentage of women involved in profession. Because professions from the neutral and male bias groups do not show significant differences, we fit a logistic regression which will predict whether a profession is in the female bias group or not. So, the neutral and male bias groups were processed together.

| | | Accuracy: 0.99 | Pseudo R-square: 0.61 |
|---|---|---|---|
| | **Coef.** | **p** | **95% conf. int.** |
| Percentage of involved women | 0.364 | 0.008 | [0.097, 0.632] |
| (Intercept) | -35.534 | 0.005 | [-60.44,-10.62] |

**Table 3:** Results of the best fitted logistic regression model with the binary dependent variable of a profession being in the female bias group. The coefficient for “Percentage of involved women” reveals that, we will see 44% increase in the odds of being in a female bias group for a one-unit increase in percentage of involved women, since exp(0.364) = 1.44. For example, probability of a profession being in the female bias group for a profession with 20% women in Germany equals 5.3e-13 and 0.55 if a profession has 98% women.

The logistic regression coefficients (Table 3) reveal that for a one-unit increase in the percentage of employed women will result in 44% increase in the odds of being in the female bias group (versus being in the male or neutral bias groups). In other words, increase in the percentage of employed women leads to probability increase of a profession being in the female bias group.

The output indicates that the percentage of employed women is significantly associated with the probability of being in the female bias group. For example, it gives the estimated probability of 0.001 for being in the female bias group for a profession with 80% employed women. The estimated probability is instead 0.63 for a profession with 99% employed women. To sum up, only professions with very high (> 97 %) percentage of women in the labor market, will be more likely to have only an article with a female title. Otherwise professions are more likely to have articles with a neutral or male title or they have articles with both male and female title.

Results of rank-sum tests and fitted logistic regression model support the hypothesis that professions with higher percentage of women have only a female page (i.e., professions from female bias redirection group). At the same time, the threshold of percentage of employed women, where one observes increase in likelihood of profession to be in the female bias, is very high (about 97% employed women).

However, we cannot identify a significant difference between the male and neutral bias groups. As can be expected, the relation between these groups cannot be described using the percentage of employed women in the labor market. Accordingly, professions which are represented only via male article, have about the same percentage of involved women as professions with both male and female articles, or are represented via a neutral article title on Wikipedia. Labor market data cannot explain and describe numerically why some professions have also female article besides male one.

Nevertheless, the difference between professions with only male and only female articles is clear. Professions, represented via articles with a male title on Wikipedia, are more likely to have a lower percentage of employed women than professions represented only via female articles. The same tendency is observed between professions, which have male and female articles or article with a neutral title (the neutral bias group), and professions with only female article titles (the female bias group).

### 5.1.3. Summary

Analysis of article titles and redirections among profession titles in the German Wikipedia reveals that most professions (812) are represented only via articles with male titles of profession. Moreover, the most encountered redirections are from female to male title. These evidences support gender disparity along article titles in the profession domain.

Additionally, we found that these choices of article titles can be explained by general popularity of male over female profession names on the German speaking World Wide Web. Hence, Wikipedia reflects the general gender bias observed on the Web. Moreover, professions, which are represented only by an article with a female title, have more sources for female profession names than male ones. Professions, which are represented only by an article with a male title, have more sources for male profession names correspondingly.

Turning to labor market statistics, we observe a significant relation between the percentage of women involved in a profession and the probability of having only a female article on Wikipedia. Thus, professions with a very high (>97%) percentage of women are more likely to be represented only by a female article. At the same time, using labor market statistics we cannot distinguish between professions, which have male and female articles simultaneously, and professions represented only by a male article.

## 5.2. Images Analysis

From the corpus of profession articles we extracted 906 images which belong to 345 profession articles. These images were classified by workers of a designated CrowdFlower task. Each image was assessed by at least three workers.

General reliability of agreement between CrowdFlower contributors were achieved with the 0.75 score of Fleiss' kappa. Overall, the Fleiss' kappa score considered to be moderate in range 0.41-0.6, substantial in range 0.61-0.8 and almost perfect agreement in range 0.81-1.00. Thus, we can assume that within our CrowdFlower task relatively high agreement was observed between contributors' responses.

Workers identified whether each photo shows people or not. According to the results of the survey, 34.3% of images do not show people on it. For every image with exactly one person (31 %), workers were asked to identify the gender. If an image showed more than one person, then workers were asked to decide whether one person is dominant (more details in Subsection 4.2) so that they could identify the gender only for that dominant person (5.8%). Otherwise, the workers were asked to identify the gender of the majority on the photo, or if the gender is not recognizable or the photo shows about the equal number of men and women, then they should choose corresponding variants. Results reveal that there were 27.3% of images with several persons without a dominant person. Aggregated results for all judges are represented in Figure 9, where "women in image" and "men in image" corresponds to one woman/man or majority of woman/man at photos with several persons without a dominant person. One can see from Figure 9 that almost half (44.8%) of the images show men, whereas only 12.4% of images show women.

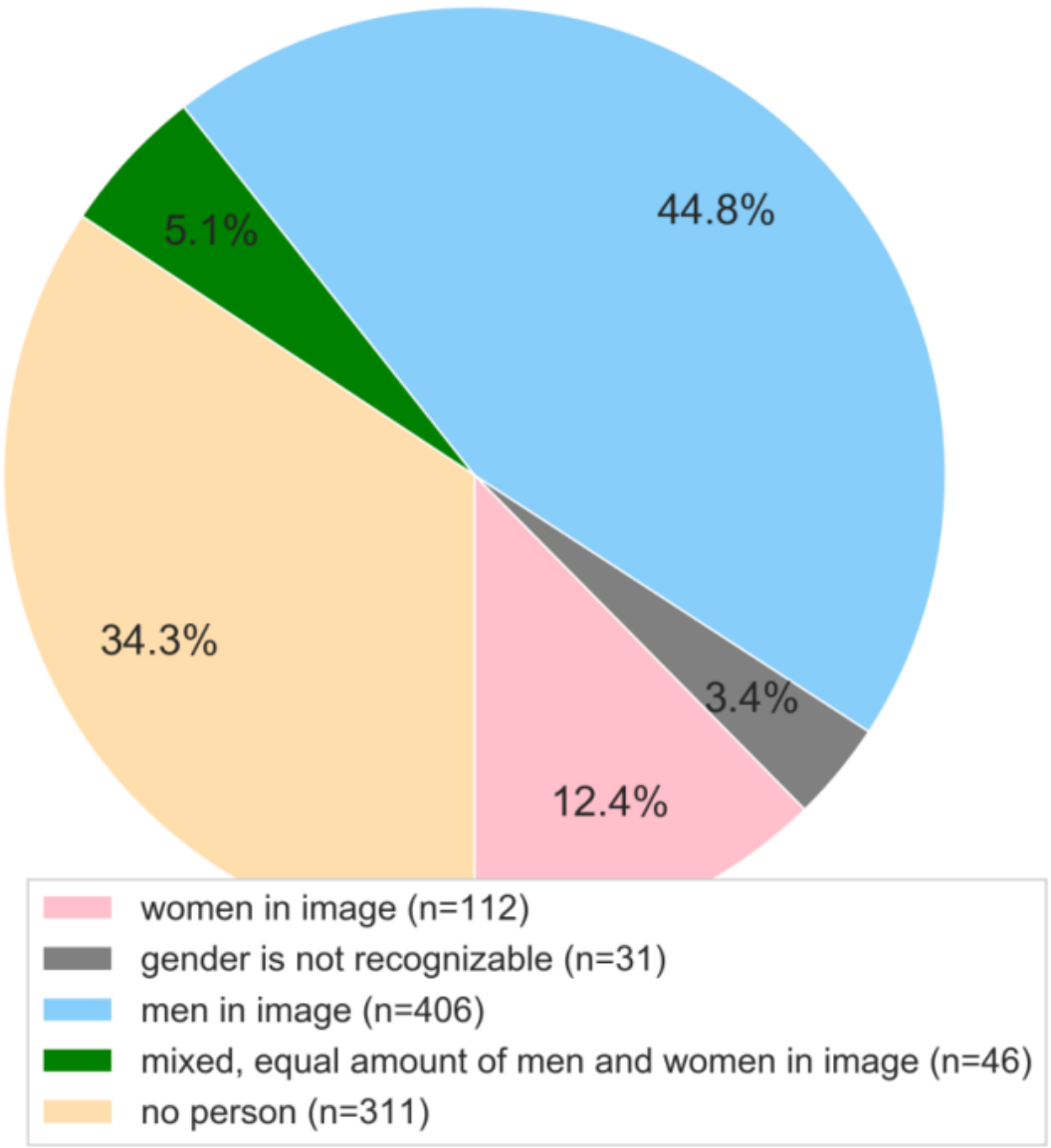

**Figure 9:** Distribution of image categories. One can see that almost four times more images depicting men than women were observed.

First, the relation was checked between the **gender of an article title** and the image categories. Thus, we focus on whether articles with female title have more images depicting women than articles with male title, and the other way around, whether articles with male title have more images depicting men than articles with female title.

In order to test the statistical independence between the image categories and the gender of article title, images were grouped according to the gender of the article title where they were observed. Thus, three groups were obtained: articles with male, neutral, and female titles. Figure 10 shows proportions of image categories, grouped by gender of profession name used as the title of the Wikipedia article.

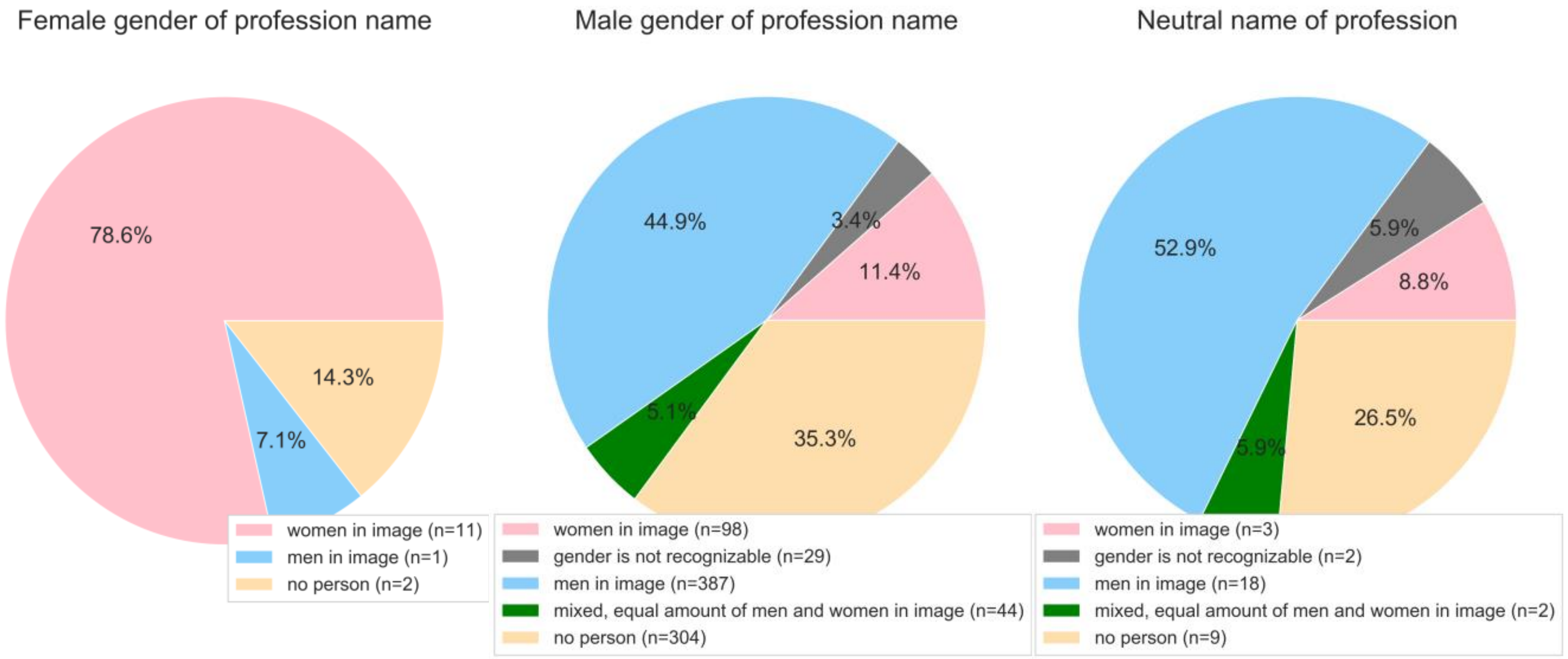

**Figure 10:** Distribution of image categories, grouped by gender of article title (profession name) where they were encountered. One can see that articles with a female title have almost ten times more images depicting women than men, whereas articles with male and neutral titles have almost four and six times fewer images depicting women than men.

Then the chi-square independence tests were performed in order to identify significant differences among the article groups. The results of chi-square tests are as follows:

a) There is a statistically significant relationship ($p<<0.0001$) between the category of images and the gender of profession titles. In other words, there is a difference in the composition of image categories between articles with female, male and neutral titles.

b) There is a significant difference between articles with male and female titles ($p<<0.0001$), and articles with female and neutral titles ($p<<0.0001$).

c) There is no significant difference between the proportion of image categories in articles with male and neutral title. Both article groups have almost 50% of images depicting men and around four and six times less images depicting women (Figure 10).

Then the post hoc tests were performed for each image category, in order to check which image categories show the observed significant differences between: i) articles with male and female titles; ii) articles with female and neutral titles. Results reveal that images depicting women ($p<<0.0001$) and images depicting men ($p<0.05$) show a significant difference between articles with male, female, and neutral titles. A significantly higher proportion of images depict women in articles with a female title than other articles. A significantly higher proportion of images depict men in articles with male titles than in articles with female titles. Interestingly enough, we also observed a significantly higher proportion of images depicting men and lower proportion of images depicting women in articles with neutral titles than in articles with female titles.

Second we tested the statistical independence between the image categories and the **redirection bias of profession** (see Subsection 5.1). So, the images were grouped according to the redirection bias of profession where images were encountered. Figure 11 shows distribution of image categories grouped by female, male, and neutral bias.

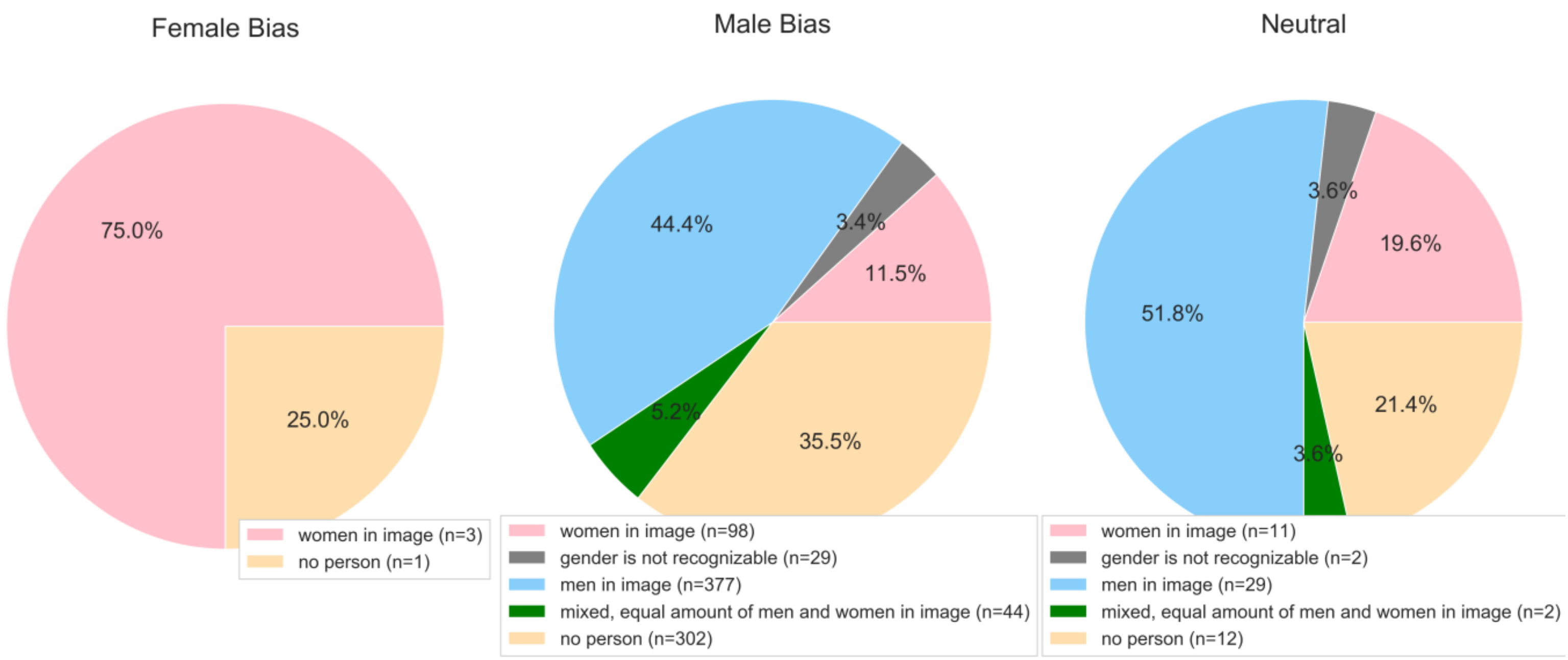

**Figure 11:** Distribution of image categories, grouped by the redirection bias of profession. One can see that female bias professions have majority of images depicting women and no images depicting men, whereas male bias and neutral professions have majority of images depicting men.

Among the professions described on Wikipedia, the proportions of image categories and the redirection bias of professions were significantly associated ($p<0.05$). According to the multiple post-hoc tests only the category of images depicting women show significant difference ($p<0.01$) between the female male bias and neutral redirection groups. One can see from Figure 11 that professions in the female bias group have only images depicting women and no images depicting men. At the same time, slightly more images depicting men and women were observed in the neutral bias group when compared to male bias group.

To sum up, the analysis of images, grouped by gender of article title (profession name) where images were encountered, reveals a significant difference in image composition. Articles with female title have almost 10 times more images depicting women than men, whereas articles with male title have 4 times more images depicting men than women. Articles with neutral title have 6 times more images depicting men than women. Thus, articles with female title and male titles both show gender inequalities. We observe evidence for female bias in articles with female titles. Basically, men are underrepresented in articles with female titles. At the same time, we observe evidence for bias against women in articles with male and neutral titles, meaning that almost 4 times more images depicting men than women were observed.

The same tendencies are encountered, when one groups images by redirection bias of profession. Namely, professions with only an article with a female title on Wikipedia have most images depicting women (75% of images), which provides evidence for female bias in these professions. Nevertheless, professions from “male bias” and “neutral bias” redirection groups are not gender equal in terms of image representations, as the majority of images depicts men (44.4% and 52% correspondingly).

It would be interesting to know how composition of images relates to labor market statistics. Thus, the following subsection represents a comparison of distributions of image categories with respect to labor market data as well as an analysis of the relation between the number of images depicting men/women and the labor market statistics.

### 5.2.1. German Labor Market

First, in order to check whether images from the profession articles reflect labor market statistics, an analysis of image groups was performed. Professions were divided in two groups: professions with female majority and professions with male majority. Thus, professions with more than 50% women were in one group and professions with more than 50% men were in another group.

Naturally, if profession images on Wikipedia reflect the German labor market statistics, one will observe the larger part of images depicting women in the group of professions with female majority and the larger part of images depicting men in the group of professions with male majority.

Figure 12 shows the distribution of image categories, grouped by gender of majority in the labor market. One can see that there are almost two times more images depicting men (38.4%) than women (20.4%) in professions with more than 50% women in labor market (female majority group). At the same time, there are almost 6 times more images depicting men (47.4%) than women (8.4%) in professions with male majority. In both groups we observe that the majority of images depict men. Therefore, we can conclude that profession images on Wikipedia do not reflect the German labor market statistics.

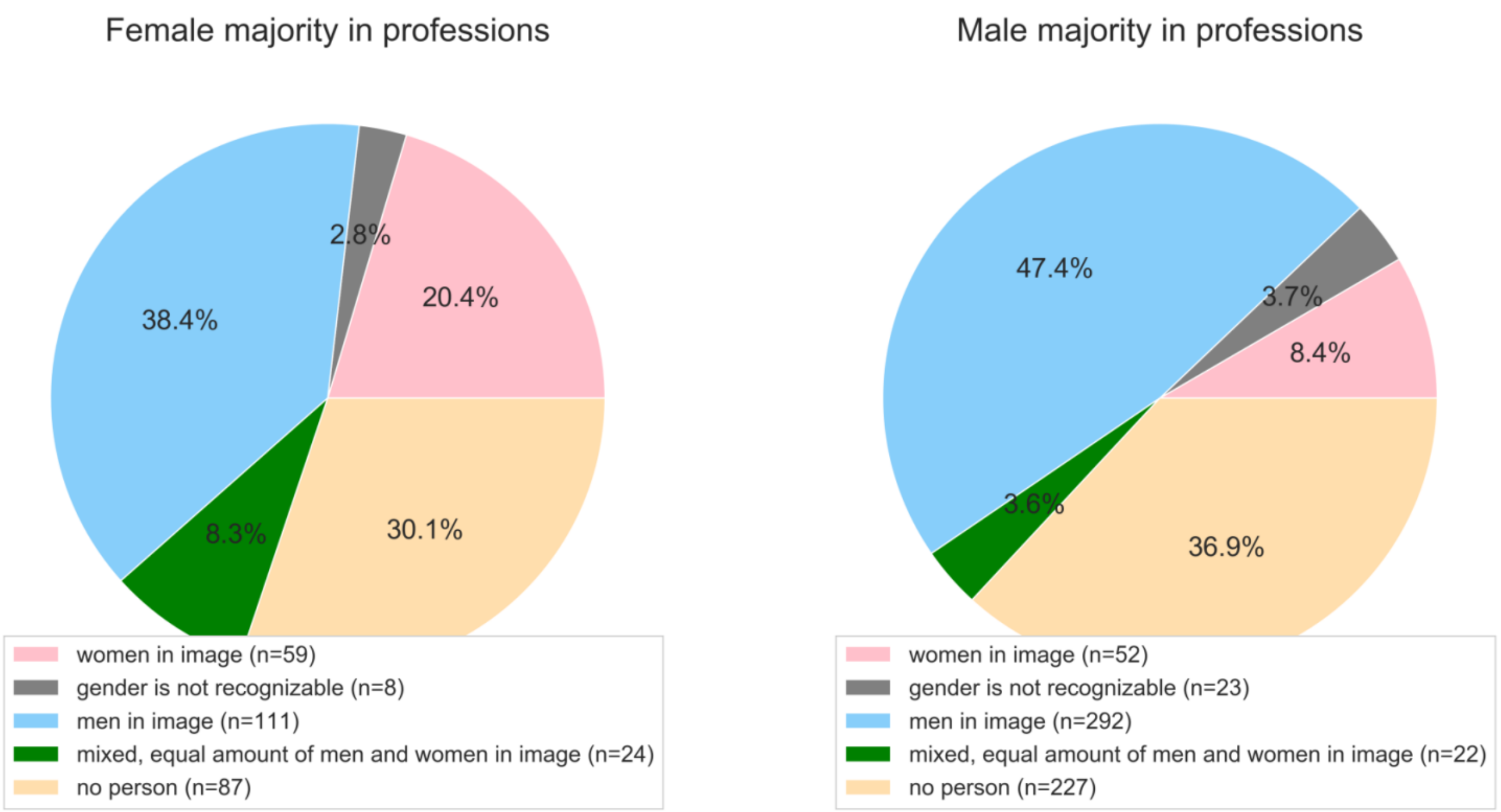

**Figure 12:** Distribution of image categories, grouped by gender of majority in the labor market. One can see that professions with female majority in labor market have almost two times more images depicting men than women, whereas professions with male majority have even fewer images depicting women, i.e., around 5.6 times more images depicting men than women.

Further analysis of these groups reveals that a statistically significant relationship ($p<<0.0001$) exists between the category of images and the gender of majority in a profession. In other words, there is a difference in the image categories composition between the female and male majority professions. According to multiple post-hoc tests, images depicting women ($p<<0.0001$), images depicting men ($p<0.01$), images with equal number of men and women ($p<0.01$), and images without people ($p<0.05$) show a significant difference between the female and male majority professions. This means that one can observe a significantly higher number of images depicting men in the group of male than female majority professions and a significantly higher number of images depicting women in the group of female than male majority professions. Nevertheless, there are almost two times more images depicting men than women in professions with female majority (Figure 12). Since, the larger part of images depicting men (38.4%) in the group of professions with female majority, we cannot state that profession images on Wikipedia reflect the German labor market statistics.

Second, it would also be interesting to observe professions with higher percentage of men and women in the labor market. Thus, we look at groups with more than 70% men and women in the labor market. Other groups can be found in the ipython notebook[12].

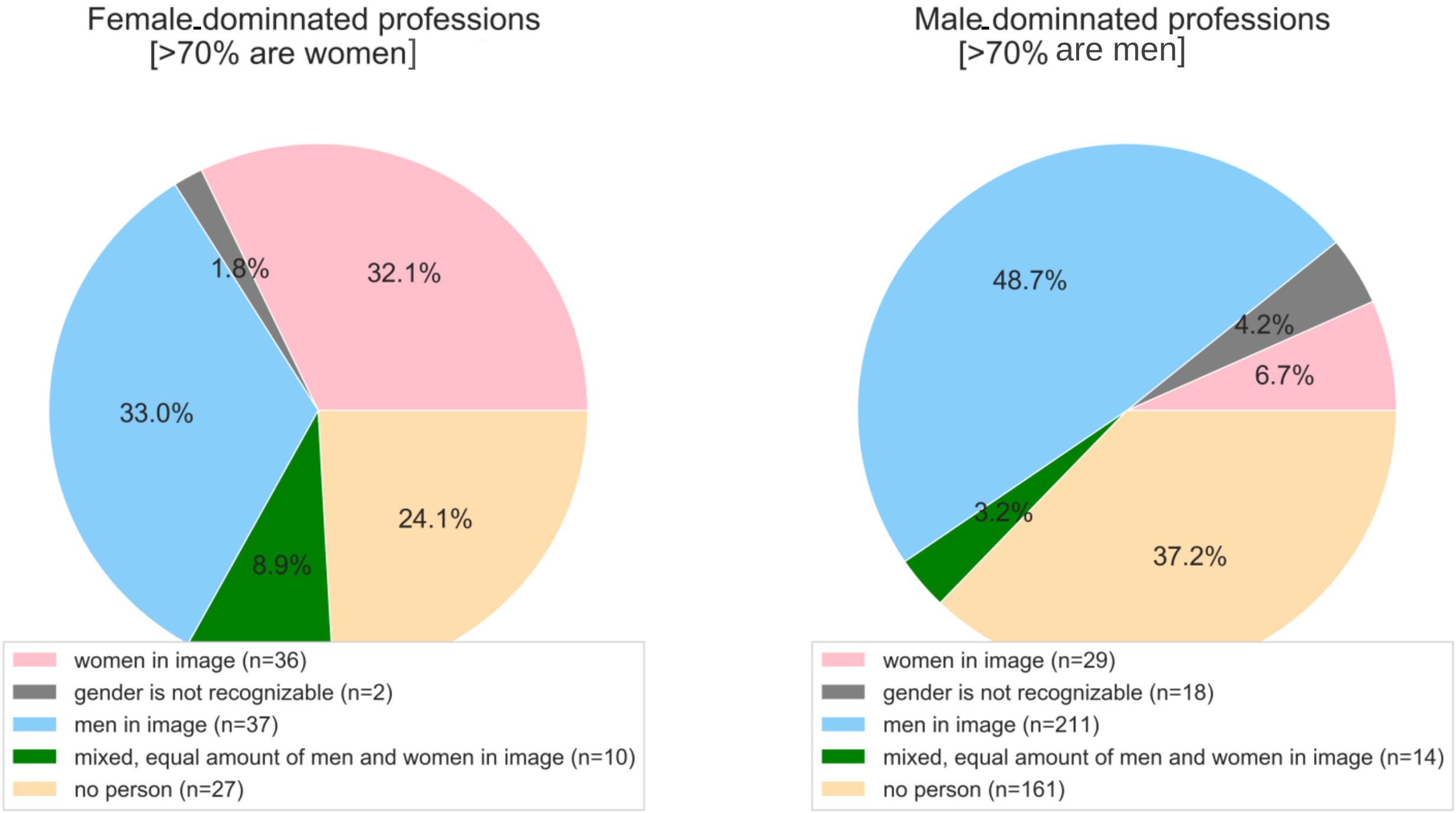

**Figure 13:** Distribution of image categories, grouped by dominating gender in the labor market. Groups consist of professions which have at least 70% men or women, respectively. One can see that professions with at least 70% women have almost equal number of images depicting men and women, whereas professions with at least 70% men have about 7 times more images depicting men than women.

[12] https://github.com/gesiscss/Wikipedia-Language-Olga-master/blob/master/9.3%20Analysis%20of%20images.ipynb

Next, professions were divided in two groups: professions with at least 70% men and at least 70% women in the labor market. Figure 13 represents the distribution of image categories among these groups. One can see that there is almost an equal number of images depicting men and women (33.0% and 32.1%) in the group of professions with at least 70% women in the labor market. In contrast, professions with at least 70% men have about 7 times more images depicting men than women (48.7% vs 6.7%).

We found a statistically significant relationship (p<<0.0001)) between the category of images and the dominant gender of the profession. In other words, a significant difference in the image categories composition was observed between the female- and male-dominated professions. According to the multiple post-hoc tests, images depicting women (p<<0.0001), images depicting men (p<0.01), images with equal number of men and women (p<0.01), and images without people (p<0.01) showed a significant difference between the female- and male-dominated professions.

An analysis of the images, grouped by gender of majority (>50%) and by dominating gender (>70%), according to the labor market statistics, reveals a significant difference in image composition. Professions, with more than 50% women in the labor market, have 2 times less images depicting women than men, whereas professions, with more than 50% men in the labor market, have almost 6 times less images depicting women than men. In other words, majority of one gender in the labor market does not imply majority of the same gender in images. The image majority remains biased against women. If professions are limited to those which have in the labor market at least 70% persons of one gender, an equal number of images will be encountered in articles of female-dominated professions and gender inequality (bias against women) in articles of male-dominated professions.

Finally, we want to know if the number of images depicting men/women relates to the labor market statistics. So, numbers of images in profession articles were correlated with the German labor market statistics for professions. Table 4 represents Spearman’s rank correlation coefficients.

| Feature 1 | Feature 2 | Correlation |
|---|---|---|
| number of images depicting women | number of women in the labor market | **0.15**$^{*}$ |
| number of images depicting men | number of men in the labor market | 0.088 |
| percentage of images depicting men | percentage of men in the labor market; | **0.3**$^{***}$ |
| percentage of images depicting men | percentage of women in the labor market | **-0.3**$^{***}$ |
| percentage of images depicting women | percentage of women in the labor market | **0.34**$^{***}$ |
| percentage of images depicting women | percentage of men in the labor market | **-0.34**$^{***}$ |
| number of images depicting women | percentage of women in the labor market | **0.34**$^{***}$ |

| Feature 1 | Feature 2 | Correlation |
|---|---|---|
| number of images depicting women | number of people in the labor market | -0.002 |
| number of images depicting women | number of men in the labor market | **-0.1**$^{.}$ |
| number of images depicting men | percentage of men in the labor market | **0.2**$^{**}$ |
| number of images depicting men | number of people in the labor market | 0.03 |
| number of images depicting men | number of women in the labor market | -0.06 |
| percentage of images depicting men | number of people in the labor market | -0.07 |
| percentage of images depicting women | number of people in the labor market | 0.01 |
| percentage of images depicting women | number of women in the labor market | **0.17**$^{**}$ |
| percentage of images depicting women | number of men in the labor market | -0.09 |

**Table 4:** Spearman’s rank correlation coefficients between the number of images in Wikipedia article about a profession and the labor market statistics of the profession.

The results reveal a moderate positive correlation (corr.coef 0.34) between the percentage of images depicting women in an article and the percentage of women in the corresponding profession, and a moderate positive correlation (corr.coef 0.3) between the percentage of images depicting men in an article and the percentage of men in the profession. Thus, professions with higher percentage of images depicting women have higher percentage of women in the labor market. Analogously professions with higher percentage of images depicting men have higher percentage of men in the labor market.

When the absolute number of images depicting women in one article was correlated with the absolute number of women in that profession, a weak positive correlation was observed (Table 4). In other words, professions with more images depicting women have more women employed in the labor market. At the same time, there is no correlation between the number of images depicting men and the number of employed men in profession. Thus, professions with high number of men can have either a lower or higher number of images depicting men.

### 5.2.2. Summary

The results of image analysis reveal that almost half (44.8%) of the images that can be found in the German Wikipedia articles about professions depict men, whereas only 12.4% of images show women. This provides evidence for gender inequalities in images in the Wikipedia profession articles. We encounter bias against women in profession images on Wikipedia. In general, women remain underrepresented on images in the articles about professions on Wikipedia.

On the one hand, analyses of images which were obtained from the articles with female title and images of professions with only a female article reveal evidence for female bias within these professions. In other words, we observe more images depicting women than men.

On the other hand, most of images depicting men in the groups of professions with female and male majority in the labor market. The same can be found for articles with neutral titles. Thus, we observe evidence for male bias in these profession groups.

Although professions with higher percentage of images depicting women have higher percentage of women in the labor market, we cannot conclude that profession images on Wikipedia reflect the German labor market statistics, since there is no correlation between the number of images depicting men and the number of men in that profession.

## 5.3. Mentioned people analysis

There are 411 articles about professions which mention at least one person. The articles mention overall 5085 persons (4272 men and 813 women). 10.4 men and 1.9 women were mentioned on average in the articles. For each article the ratio of mentioned men was estimated; 0.83 is the mean ratio of men per article and 0.98 is the median ratio (Figure 14). In other words, out of all people mentioned in an article, 83% of them are men and only 17% of them are women on average.

Figure 14 shows the distribution of men ratios. As one can see, ½ of articles have a ratio of mentioned men close to 1.0 and ¾ of articles have a ratio of mentioned men from 0.8 to 1.0. Moreover, Figure 14 features the following examples: article “Konstrukteur” mentions 30 persons and all of them are men, article “Journalist” mentions 41 men and 18 women, article “Model” mentions 39 men and 78 women.

Since articles with only one (or two) mentioned men will have ratio of mentioned men equal 1.0, one might assume that ratio of mentioned men is a misleading measure. In order to control that effect, the relations were checked between the ratio of mentioned men and the following: i) number of mentioned men, ii) number of mentioned women. (Alternatively, we could also remove from an analysis all articles with very few mentioned persons.) A positive weak correlation (Pearson cor.coef. 0.11) was found between the ratio of mentioned men and the number of mentioned men, and a negative correlation (Pearson cor.coef. - 0.29) was found between the ratio of mentioned men and the number of mentioned women. In other words, articles that mention more men have a slightly higher ratio of mentioned men and articles with higher ratio of men mention a slightly lower number of women.

Articles which mention an equal number of men and women (i.e., the ratio of mentioned men equals 0.5[13]) are considered gender neutral. Consequently, all articles with ratios of mentioned men higher than 0.5[13] are male biased (i.e., biased against women) and all articles with the ratio smaller than 0.5[13] are female biased (i.e., biased against men). One can see from Figure 14 that most articles have a ratio higher than 0.5. In other words, male bias is observed in most articles. In particular, 92.5% of the articles are male-biased, 3.1% of the articles are female-biased, and 4.4% of the articles are gender-equal.

[13] We used the interval 0.5±0.05 as an approximation for ‘almost equal number of men and women’. Articles with a ratio of mentioned men higher than 0.55 are male biased; articles with a ratio of mentioned men smaller than 0.45 are female biased. Thus, articles which mention 5 women and 6 men will be considered as gender neutral. If articles mention less than 10 persons, only ratios of strictly equal numbers of men and women will be considered as gender equal ratios.

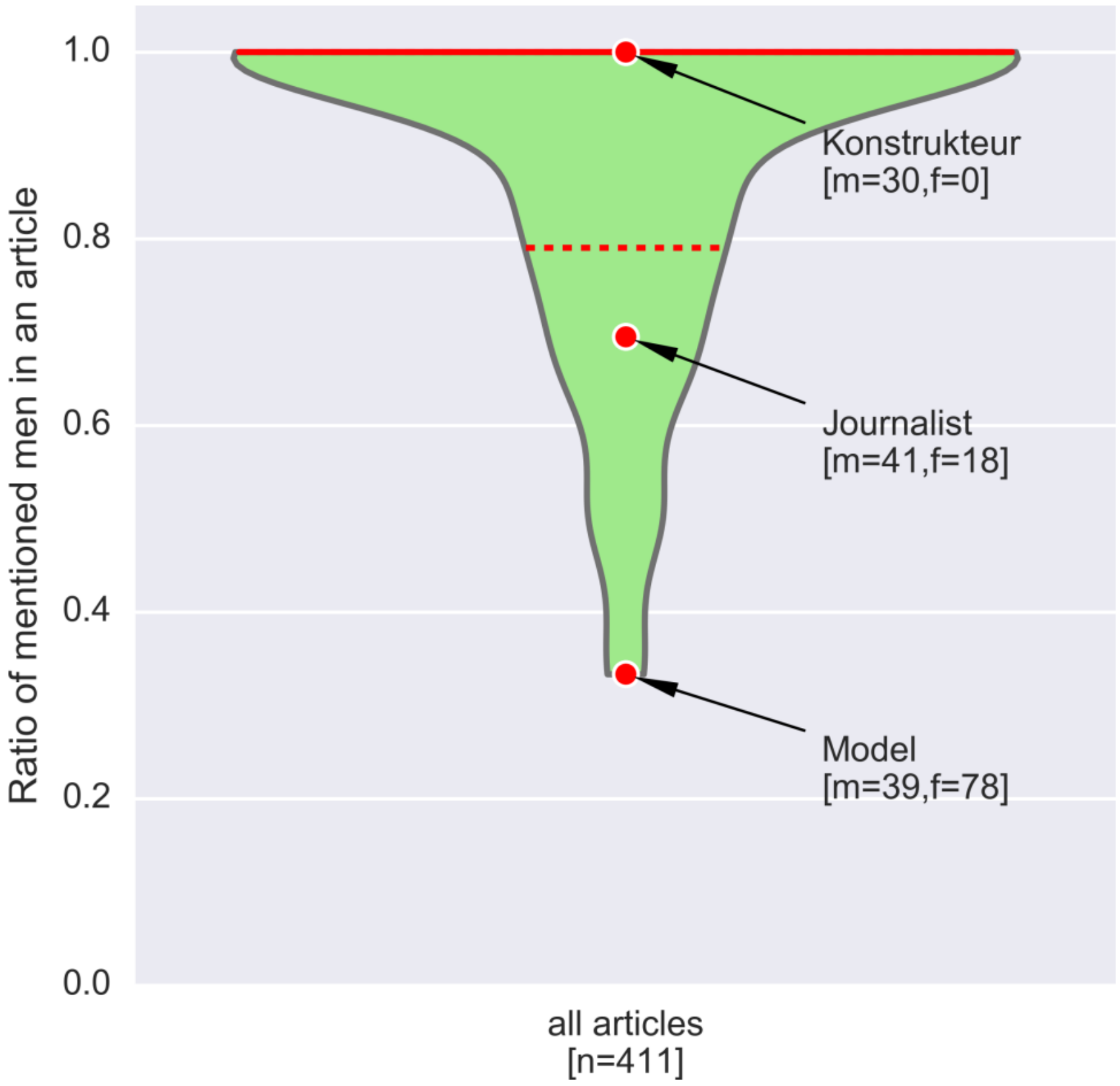

**Figure 14:** Violin plot of the distribution of mentioned men ratios in articles. The violin plot features a kernel density estimation of the underlying distribution. The shape of the violin plot represents the density: the wider the violin plot, the greater the number of articles with a particular ratio of men. The solid red line represents the median; the doted lines show 1st and 3rd quartile. The violin plot represents data points which are distributed within two standard deviations of the mean, i.e., data points which assumed to be outliers are removed. One can see that ½ of articles (according to the median value) mention only men in the articles, ¾ of articles mention 80-100% men and only 0-20% women. Moreover, only a few articles mention either the equal number of men and women or more women than men. For example, the article “Model” mentions 39 men and 78 women.

Next, the dependency was checked between the ratio of mentioned men and the **gender of the article title** where these men were mentioned. Basically, we want to know if: i) articles with a male title have a higher ratio of mentioned men than articles with a neutral or a female title; ii) articles with a neutral title have a higher ratio of mentioned men than articles with a female title.

First, the articles were grouped according to gender of article title (Figure 15). Then three rank-sum tests were performed in order to compare ratios of mentioned men of the groups. Results of the tests reveal that there is a significant difference between distribution of values in the articles with male versus female title ($p<0.05$). Ratios of mentioned men in articles with male title (median is 0.98) are more likely to be larger than the ratios in articles with female title (median is 0.65) (Figure 15). There is no significant difference between pairs of groups: male-neutral, female-neutral.

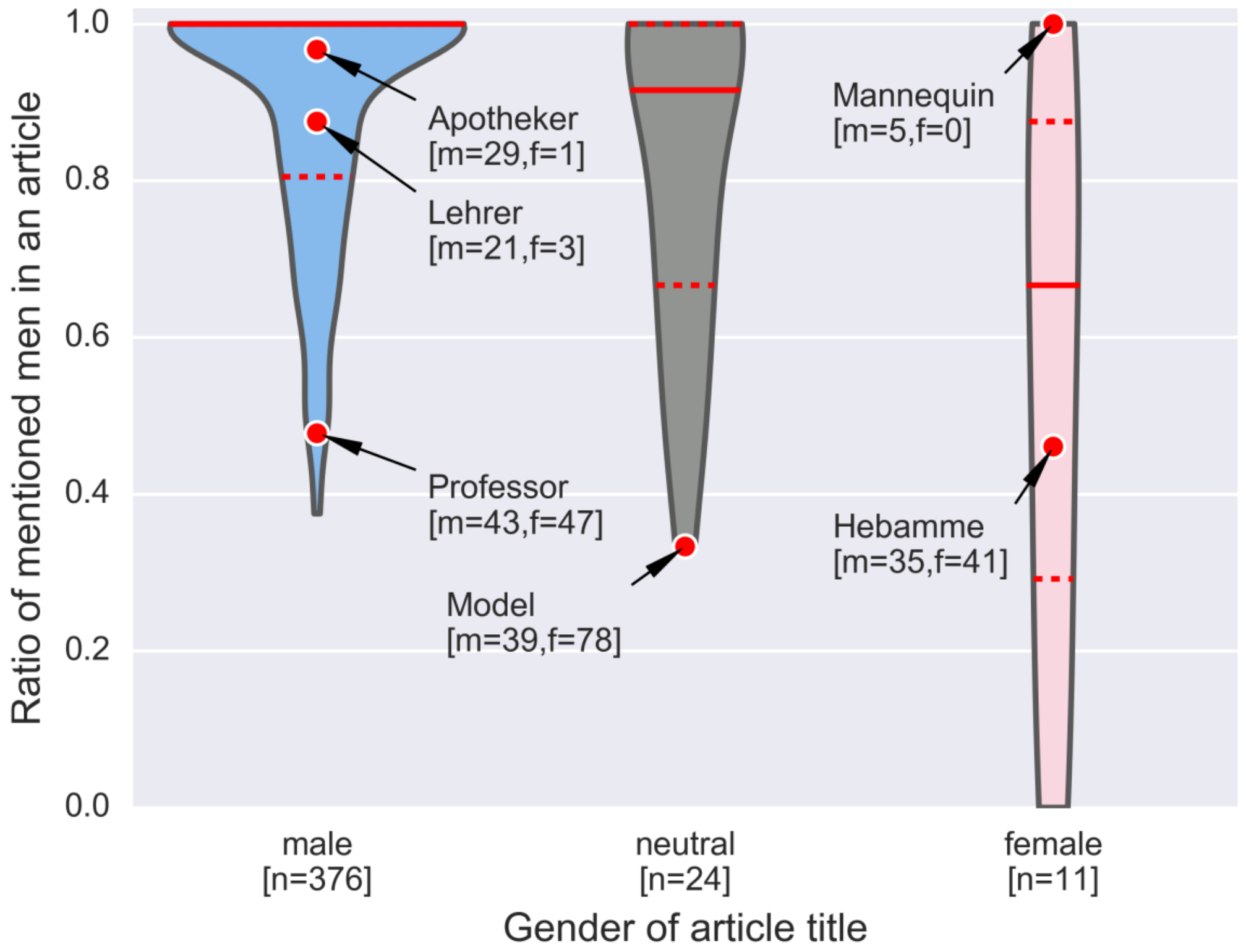

**Figure 15:** Violin plot of the distribution of mentioned men ratios in articles, grouped by gender of profession name. The violin plot features a kernel density estimation of the underlying distribution. The shape of the violin plot represents the density: the wider the violin plot, the greater the number of articles with a particular ratio of men. The solid red line represents the median; the doted lines show 1st and 3rd quartile. The violin plot represents data points which are distributed within two standard deviations of the mean, i.e., data points which assumed to be outliers are removed. One can see that articles with female title overall mention from 0 to 100% men and ½ of the articles mention 65-100% men and 0-35% women. Nevertheless, a few articles with female title mention more men than women.

Although, a significantly lower ratio of mentioned men was observed in articles with female title than in articles with male title, the median ratio of mentioned men in articles with female title equals 0.65. Other groups had even higher medians, i.e., 0.98 and 0.83 in the articles with male title and neutral title correspondingly. Thus, there are more mentioned men than women (i.e., ratio of mentioned men > 0.5) in the majority of the articles from all three groups. This implies that for all groups we observed mostly male bias with few exceptions in each group, where we observed female bias. Examples of the exceptions follow: “Gleichstellungsbeauftragte”, “Model” (Figure 15). Examples of articles that mention almost equal number of men and women follow: “Hebamme” and “Professor”.

Then, the significant differences in ratios of mentioned men were checked between professions which have only a male article (“male bias” redirection group), only a female article (“female bias” redirection group) and professions with either articles of both genders or a neutral article (“neutral bias” redirection group). The rank-sum tests reveal that there is no significant differences ($p>0.05$) in ratios of mentioned men between these groups of professions.

To summarize, the results of the mentioned people analysis reveal that ½ of articles have a ratio of mentioned men between 0.8 and 1. In other words, ½ of articles mention 80-100% of the time men and only 0-20% of the time women. This provides evidence for gender inequalities, meaning that most articles about professions mention more men than women. Even though, articles with female title have significantly lower ratio of mentioned men than articles with male title, most articles in both groups have ratios of mentioned men higher than 0.5, thereby providing evidence for gender disparity.

### 5.3.1. German Labor Market

Next the relation was checked between the labor market statistics of a profession and the number of mentioned men/women in the profession article on Wikipedia. To begin, the difference in ratios of mentioned men was checked between professions with male and female majority. Then we check the relations between the number of mentioned men/women and the number of employed men/women in a profession.

In order to check differences in ratios of mentioned men, the professions were divided in two groups: i) professions with more than 50% employed women; ii) professions with more than 50% employed men. Figure 16 shows the distributions of mentioned men ratios in articles with male and female majority in the German labor market. Then the rank-sum test was performed between values of these two groups.

The analysis reveals a significant difference between the distributions of men ratios in professions with male and female majority ($p<<0.0001$). The ratios of mentioned men in the articles with male majority (median ratio is 0.98) are more likely to be larger than the ratios of mentioned men in the articles with female majority (median ratio is 0.88). In other words, professions with female majority have articles with a significantly lower ratio of mentioned men than articles of professions with male majority.

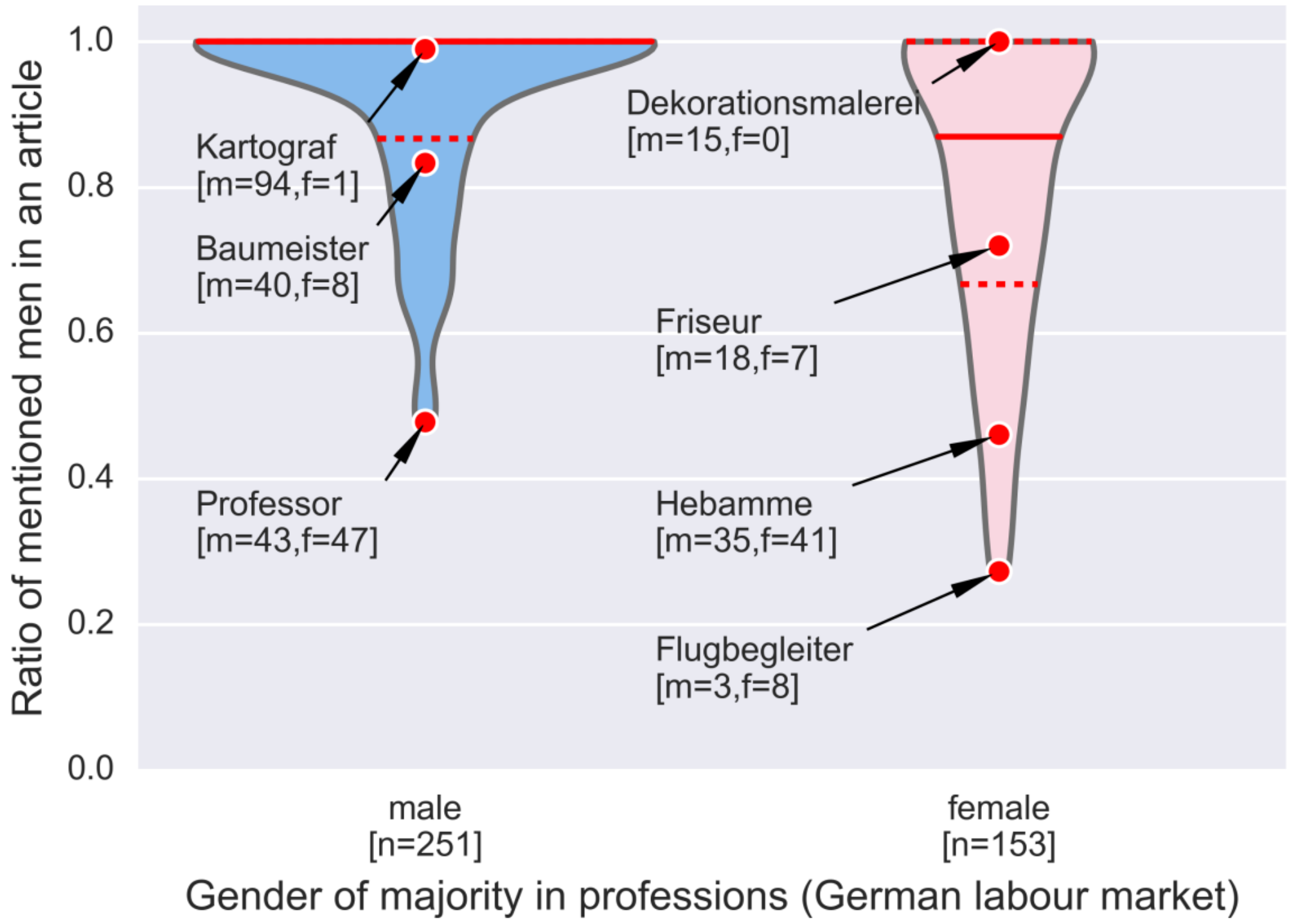

**Figure 16:** Violin plot of the distribution of mentioned men ratios in articles, grouped by gender of profession majority. The violin plot features a kernel density estimation of the underlying distribution. The shape of the violin plot represents the density: the wider the violin plot, the greater the number of articles with particular ratio of men. The solid red line represents the median; the doted lines show 1$^{st}$ and 3$^{rd}$ quartile. The violin plot represents data points which are distributed within two standard deviations of the mean, i.e., data points which assumed to be outliers are removed. One can see that professions with female majority have ¾ of articles that mention at least 68% men and at most 32% women, whereas professions with male majority have ¾ of articles with at least 88% mentioned men and at most 12% mentioned women.

Nevertheless, ¾ of articles (Figure 16), among professions with female majority in the labor market, mention 68%-100% men and 0-32% women. This implies that the larger part of the articles for professions with female majority expose male bias, despite the fact that professions with female majority have significantly lower ratios of mentioned men in the article.

Next we want to know the strength of the relation between the percentage of mentioned men/women in the article and the percentage of employed men/women in the profession. Results of Spearman’s rank correlation (Table 5) reveal weak positive correlation (corr. coef.0.27) between the percentage of mentioned women and the percentage of women in the labor market. The same correlation can be found for the percentage of mentioned men and the percentage of men in the labor market. In other words, the higher is the percentage of women in a profession, the higher is the percentage of mentioned women in the article about the profession, and the other way around. In summary, one can conclude that the percentage of mentioned men or women reflects the German labor market statistics in terms of gender proportions.

| Feature 1 | Feature 2 | Correlation |
|---|---|---|
| percentage of mentioned women | percentage of women in the labor market | **$0.27^{***}$** |
| percentage of mentioned men | percentage of men in the labor market | **$0.27^{***}$** |
| number of mentioned men | number of men in the labor market | **$-0.23^{***}$** |
| number of mentioned men | number of women in the labor market | **$-0.15^{**}$** |
| number of mentioned men | number of people in the labor market | **$-0.20^{***}$** |
| number of mentioned women | number of men in the labor market | -0.08 |
| number of mentioned women | number of women in the labor market | $0.09^{.}$ |
| number of mentioned women | number of people in the labor market | -0.01 |
| number of mentioned persons | number of men in the labor market | **$-0.21^{***}$** |
| number of mentioned persons | number of women in the labor market | **$-0.09^{.}$** |
| number of mentioned persons | number of people in the labor market | **$-0.17^{***}$** |
| number of mentioned women | percentage of women in the labor market | **$0.26^{***}$** |
| number of mentioned men | percentage of men in the labor market | **$-0.10^{*}$** |
| percentage of mentioned women | number of women in the labor market | **$0.17^{***}$** |

**Table 5:** Spearman's rank correlation coefficients between the number of mentioned men/women in a Wikipedia article about a profession and the labor market statistics of the profession.

We also check whether correlation changes when absolute numbers rather than percentages are considered. Thus, the absolute numbers of mentioned men/women in the article about profession were correlated with the absolute numbers of men/women in the profession. The results are as follows:

a) weak negative correlation (corr.coef. -0.23) between the number of mentioned men in the article and the number of employed men in the profession. In other words, the more men are working in the profession, the fewer men are mentioned in the article.
b) weak negative correlation (corr.coef. -0.15) between the number of mentioned men in the article and the number of employed women in the profession. In other words, the more women are working in the labor market, the fewer men are mentioned in the article.
c) weak negative correlation (corr.coef. -0.20) between the number of mentioned men in the article and the number of employed people in the profession. In other words, the more people are employed in the profession, the fewer men are mentioned in the article.
d) no correlation between the number of mentioned women in the article and the number of employed women in the profession;

e) no correlation between the number of mentioned women in the article and the number of employed men in the profession;
f) no correlation between the number of mentioned women in the article and the number of employed people in the profession.

Other correlation results are presented in Table 5.

Interestingly, professions, with fewer employed people (and men), mention more men in the articles about these professions than those professions which have more employed people (and men). In other words, the less popular professions in the German labor market mention more men than the popular professions.

One explanation of that phenomenon might be that editors prefer to write about historical development of professions and prefer to mention prominent historical persons in Wikipedia articles, such that some professions are described in a way that they cover many historical persons in these articles.

In order to control the historical phenomena, the mentioned people were limited to those who reflect current labor force. Thus, the pool of people was restricted to those who were born after 1960. (In our study we also checked other thresholds; results are available in ipython notebook[14].) Figure 17 represents distributions of ratios of mentioned men in articles; the right plot represents the ratios of men among people who were born before 1960; the left plot deals with people who were born after 1960 instead. One can see from Figure 17 that the right plot is more “peacky” near the median value of 1.0, whereas the left plot is more flat. In other words, if one takes into account only people who were born before 1960, one will find that articles mention at least 50% men and ¾ of articles mention at least 92% men. Thus implies that articles mention predominantly more men than women, if one considers only people who were born before 1960. One can see from the right plot of Figure 17 approximately ¾ of articles with at least 50% mentioned men and ½ of articles with 100% mentioned men. In other words, if one considers only people who were born after 1960, one will observe ¾ of articles with at least 50% mentioned men, i.e. articles with gender equality and male-biased articles. Taking into account percentage of articles with equal number of men and women, 71.6% of articles were revealed as male-biased (i.e., ratio of mentioned men >0.5[13]), 6.8% of articles were revealed as gender-equal (i.e., ratio of mentioned men equals 0.5[13]) and 21.9% of articles were revealed as female-biased (i.e., ratio of mentioned men <0.5[13]).

The comparison between ratios of mentioned men among *all mentioned people* (Figure 14) versus *those who were born after 1960* (Figure 17, right plot) revealed a significant decrease in ratios of mentioned men (p<<0.001). In particular, ¾ of articles were observed with at least 79% men and at most 21% women (Figure 14) if one considers all mentioned people in an article, whereas, in the second case (Figure 17, the right plot), ¾ of articles were observed with at least 50% men and at most 50% women.

[14] https://github.com/gesiscss/Wikipedia-Language-Olga-master/blob/master/9.2.1%20Analysis%20of%20mentioned%20people%20(restricted%20by%20BirthDate).ipynb

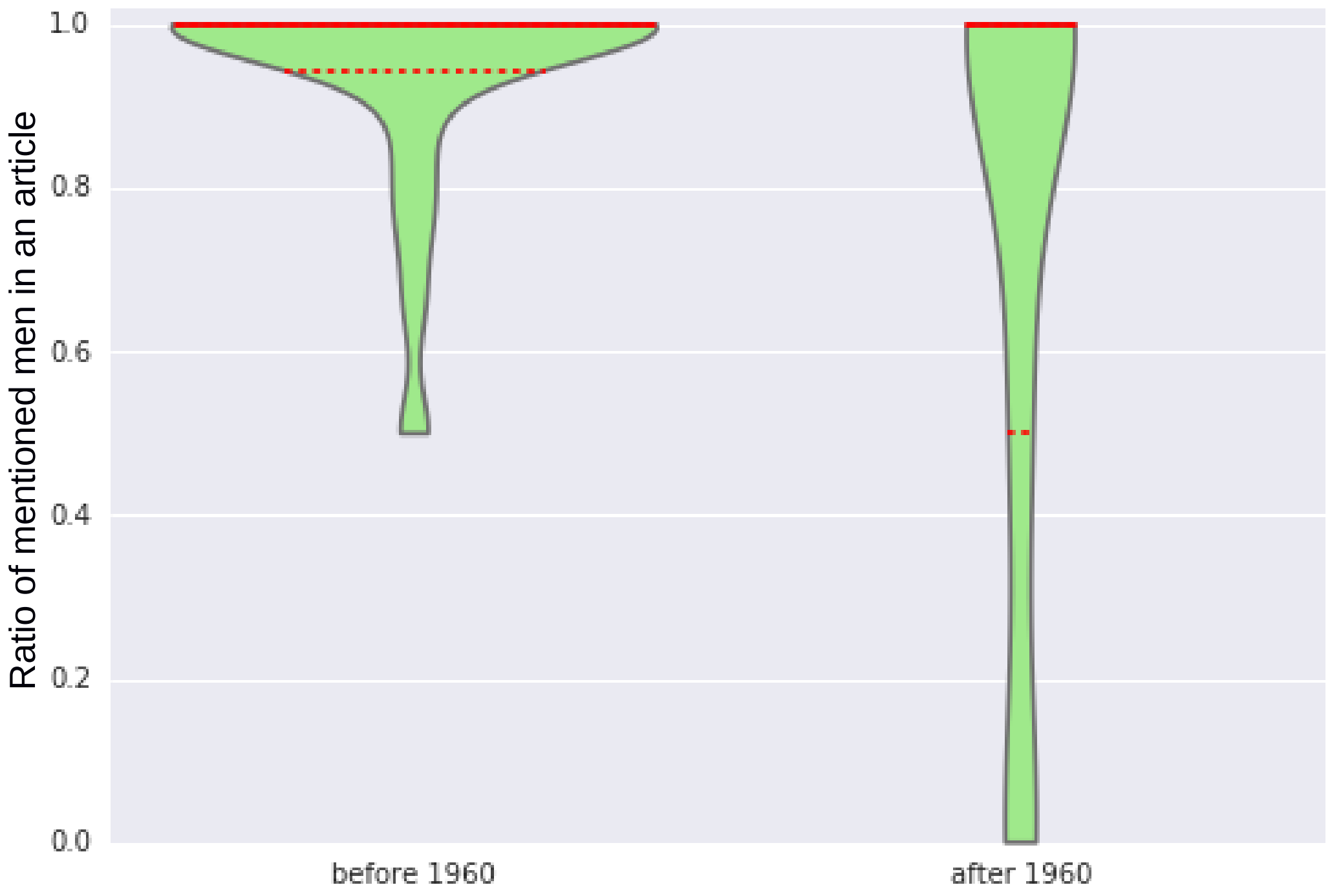

**Figure 17:** Distributions of ratios of mentioned men in articles, the **right plot** represents ratios of men among people who were born before 1960; the **left plot** deals with people who were born after 1960 instead. According to the rank-sum test, the difference between these distributions is significant ($z = -12.6$, $p << 0.001$). One can see that ¾ of articles mention at least 94% men if one considers only people who were born before 1960, whereas ¾ of articles mention at least 50% men in the second case.

Next we check whether the negative correlations persist if one considers separately people who were born before and after 1960. Table 6 and Table 7 represent Spearman's rank correlation coefficients, if one considers only people who were born before 1960 and after 1960. As one can see, the correlation coefficients decrease (compare to the case when all persons are considered, Table 5) between the number of mentioned men and the number of employed people/men/women for case "after 1960". In other words, the strength of association, between the number of mentioned men and the number of employed people/men/women, decreases.

| Feature 1 | Feature 2 | Correlation |
|---|---|---|
| percentage of mentioned women | percentage of women in the labor market | **0.18**$^{**}$ |
| percentage of mentioned men | percentage of men in the labor market | **0.18**$^{**}$ |
| number of mentioned men | number of men in the labor market | **-0.20**$^{**}$ |
| number of mentioned men | number of women in the labor market | **-0.15**$^{*}$ |
| number of mentioned men | number of people in the labor market | **-0.19**$^{**}$ |

**Table 6:** Spearman's rank correlation coefficients when only people born before 1960 were considered; correlations between the number of mentioned men/women and the labor market statistics of profession.

| Feature 1 | Feature 2 | Correlation |
|---|---|---|
| percentage of mentioned women | percentage of women in the labor market | **0.23**$^{.}$ |
| percentage of mentioned men | percentage of men in the labor market | **0.23**$^{.}$ |
| number of mentioned men | number of men in the labor market | -0.12 |
| number of mentioned men | number of women in the labor market | -0.11 |
| number of mentioned men | number of people in the labor market | **-0.12**$^{*}$ |

**Table 7:** Spearman's rank correlation coefficients when only people born after 1960 were considered; correlations between the number of mentioned men/women and the labor market statistics of profession.

To conclude, if one considers only people born after 1960, the strength of negative correlation, between the number of mentioned men and the number of employed people, decreases. This also implies that the negative correlation is observed between the number of mentioned men and the number of men/women/people can be found for fewer articles if only people born after 1960 are considered.

### 5.3.2. Summary

More mentions of men than women were encountered in most articles. In particular, out of all people mentioned in an article, 83% of them are men and only 17% of them are women on average. Besides, 92.5% of the articles are male-biased, i.e., more than 50% of mentioned people are men. However, a few articles had either equal number of mentioned men and women (4.4 % of articles) or more mentioned women than men (3.1% of articles). In other words, gender equality and female bias were observed in a few articles.

The analysis of articles with female title and articles of professions with female majority in the labor market reveal significantly lower ratios of mentioned men. Nevertheless, even for groups where the median difference of ratios was significant, the ratio of mentioned men remains above 0.5 for at least ½ of articles, which provides evidence for male bias prevalence even in these profession groups. For example, the group of professions with female titles of articles mention 65% men (median value) and the group of professions, with women majority in the labor market, mentions 88% men (median value).

If one restricts mentioned people to those who were born after 1960, one will observe the following: at least 50% men in ¾ of articles; 100% men in ½ of articles. Considering percentages of articles, male bias (i.e., ratio of mentioned men >0.5) was observed in 71.6% of articles, gender equality (i.e., ratio of mentioned men equals 0.5) was observed in 6.8% of articles, and female bias (i.e., ratio of mentioned men <0.5) was observed in almost 21.9% of articles. Thus, we can conclude that elimination of people, who were born before 1960, leads to less skewed gender inequality towards men. However, male bias remains in almost 72% of articles.

## 6. Discussion

We observe prevalence of male bias along all three dimensions. For most professions one can only find an article in the German Wikipedia that has the male job title as the article title. Moreover, almost four times more images depicting men than women were encountered in the profession articles. Along the mentioned people dimension, male bias was observed in 92.5% of articles.

However, results of the analysis along these three dimensions exhibit female bias for a few professions. For example, along the mentioned people dimension, female bias was encountered in 3.1% of articles and in 22% of articles if only people born after 1960 are considered. Moreover, prevalence of images depicting women (i.e., female bias along the images dimension) was revealed in the following groups: articles with female titles (with or without corresponding article with male title).

**Why do gender biases exist in Wikipedia articles?** There are several possible reasons. We list possible reasons but the goal of this work was not to find a causal link between bias and cause, but to measure bias. First of all, each individual has implicit gender stereotypes and biases [49]. For instance, a person may explicitly believe that men and women are equally appropriate for professions. Nevertheless, that person may have implicit associations regarding women and home [41], and these implicit associations may influence this person's behavior in any number of biased ways, starting from preference in hiring equally qualified men over women, ending with trusting less feedback from female colleagues [9]. Moreover, some professions are implicitly associated with genders [48]. For example, nurses are implicitly stereotyped to be female and programmers are stereotyped to be male. As a result, a Wikipedia collaborator might give implicitly preference to adding image depicting a man rather than a woman when editing or writing articles for particular professions.

Why does male bias exist and predominate on Wikipedia? Previous research [5, 6, 38] reveals that a dominant majority of editors in Wikipedia are male. Hill and Shaw [5] reported 22.7% US female editors and 16.1% overall female editors on Wikipedia, which provides evidence for gender inequality in the Wikipedia editors community. As a result, Wikipedia was criticized for having fewer and less extensive articles about women or topics important to women [34].

Another reason, which potentially explains the male bias on Wikipedia, is the male bias in other media (e.g., the Web). Wikipedia might reflect and in some cases inherit biases from other media. We observe evidence for male bias on the German speaking web, i.e., for most of professions one can find much more sources for male than female professions, which might be also reflected on Wikipedia. Moreover, several studies [35, 36, 37, 51, 52, 53] reveal gender bias and stereotypes in mass media and interviews of sport competitions. Kay et al. [3] revealed the gender bias present in image search results for professions and occupations.

**What can be done to reduce biases?** Over the last five years, the Wikimedia Foundation has made many attempts to attract more female editors [44, 45]. Even though the Wikimedia Foundation was not reporting about success or failure of their attempts, attraction of more female editors is still needed.

On the other hand, not only female editors can change the article content and make it more gender neutral. If one would have transparent guidelines and rules towards gender equality in article content, the gender disparity of individual articles might decline. Therefore, Wikipedia equality rules related to professions and gender representation in the images and mentioned people scope should be developed. Following these rules, an automated warning system, applied before acceptance of revisions, could potentially improve equality in the content of articles.

**Implications.** We proposed a viable method for the identification and assessment of gender bias related to professions in Wikipedia articles. The presented method can be implemented and incorporated into a software tool helping Wikipedia editors to identify and warn about existing gender inequality in articles along three dimensions: gender-inclusiveness in job titles and corresponding redirection analysis, images analysis, and mentioned people analysis.

**Limitations and Future Work**. The proposed approach relies on human annotators (CrowdFlower contributors) for gender classification of images of people, since state-of-the-art gender classifiers of human images perform well only under certain conditions. However, our method could benefit from automated gender identification approaches by converting our asynchronous image analysis into a synchronous one.

Due to the fact that, in the past, women had a secondary role in science, culture, and history, we may observe much less women in professional domains. Thus, besides filtering out people by their date of birth, we could extract paragraphs from the text that refer to history and historical development of professions. We could perform additional analysis for the extracted text and the remaining text.

For the mentioned people analysis, all mentioned people from the articles were used. We do not filter out the mentioned people who do not have that profession. From our point of view, people who are mentioned in the profession articles are related to it, even if they do not have that profession, thus removal of persons from analysis is arguably not necessary. One could think of all mentioned persons as a suite of features which create the full and complete image about profession.

If we would aim to filter out persons, the simplest way would be to let CrowdFlower workers annotate people as to whether they have the profession or not. The further analysis would be applied to the people of the profession. A more sophisticated way would be to look at people who do not have the profession and check in what relation they stand to other mentioned people (especially those who are employed as) or basically, why these people are mentioned? This kind of task can be performed by CrowdFlower workers as well. Further, we could extend our research with gender analysis of most prominent semantic relations between mentioned people in an article. For example, we could find if people who are managers/bosses of other mentioned persons are more likely to be men.

The method proposed in our research can be applied for the analysis of gender inequalities in different Wikipedia editions. Our images and mentioned people analysis methods can be applied to any language, whereas the redirection analysis has some limitations. It can only be applied to languages with masculine and feminine grammatical genders. Moreover, further cross-language analysis can be performed, such that one could identify the least and most biased Wikipedia editions. This analysis could be interesting since the results can be compared to the Global Gender Gap Index [8] for each country associated with Wikipedia edition.

One could argue that the compiled list of professions and occupations is not complete, thus we might not be able to find all profession articles. The list consists of 4457 professions. We were able to match 885 Wikipedia articles with corresponding professions. Perhaps surprisingly, the ratio of matched articles to professions is about 20%. However, the list of profession names has too specific profession names (e.g., “Betriebsbeauftragter im Umweltschutz”) which, in some cases, have very close profession names (or even synonyms) in the same list. These profession names might be represented only by one Wikipedia article (e.g., “Abfallbeauftragter”). Thus, some of the professions from the list are implicitly represented on Wikipedia too. We found another dataset of profession names[15] which has about 27000 entries. This dataset has even more variations on job titles, for example, the profession sport teacher (without particular sport specialization) has more than 15 different names which are distinguished by job places and by degrees (e.g., “Sportlehrer/in - Sportvereine”, “Sportlehrer/in (staatl.gepr.)”, “Dipl.-Sportlehrer/in (Uni)”). A bit more than 1100 articles were found by matching the new list of profession names with Wikipedia articles. We were able to find around 20% more articles; among these articles, there were only 25 articles with female title (compare to 18 articles from the list used in this research). One might apply the proposed approach for gender inequality assessment to the new dataset.

Additionally, one could perform timestamp analysis of Wikipedia articles’ revisions. Since Wikipedia articles evolve over time, analysis of Wikipedia articles history might lead to valuable insights.

[15] Retrieved from http://statistik.arbeitsagentur.de/Statischer-Content/Grundlagen/Klassifikation-der-Berufe/KldB2010/Systematik-Verzeichnisse/Generische-Publikationen/Systematisches-Verzeichnis-Berufsbenennung.xls on 19.01.2016

## 7. Conclusion

We identify and characterize gender inequalities in the German Wikipedia articles about professions along three dimensions: gender-inclusiveness in job titles and corresponding redirections, images, and people mentioned in the articles. We find evidence for both, systematic overrepresentation of men along all three dimensions and female bias for a few professions.

Analysis of article titles and redirections on Wikipedia reveals that most professions are represented only via an article with male title of profession. Moreover, most encountered redirections are from female to male title. These evidences support gender disparity along article titles in profession domain.

Additionally, we found that choices of article titles can be explained by general popularity of male over female profession names on the German speaking Web. In other words, Wikipedia reflects the general gender bias observed on the Web. Professions represented only by an article with female title, are more likely to have more sources for female than male job titles according to the number of Google results for the title. Professions, which are represented only by articles with male titles, are more likely to have more sources for male job titles.

Turning to the labor market statistics, we observe a relation between the percentage of women involved in a profession and the probability of having only a female article on Wikipedia. Thus, professions with a very high percentage of women are more likely to be represented only by a female article. However, using labor market statistics, we cannot distinguish between professions with both male and female articles versus professions represented only by male article.

We find that almost half of images from the profession articles depict men and only around 12% of images depict women. Thus, we encounter male bias in profession images on Wikipedia. On the one hand, we observe a significantly higher number of the images depicting men in the groups of professions with female majority and male majority in the labor market. . Hence, we encounter evidence for male bias within these profession groups. On the other hand, we observe a significantly higher number of images depicting women than men in the articles with female title and professions which have only a female article on Wikipedia. Hence, we encounter evidence for female bias within these professions with respect to images depicting women.

Generally, more men than women were mentioned in most articles. Only a few articles mention either (near to) equal number of men and women or more women than men. Particularly, out of all mentioned people, we observed ¾ of articles with 79-100% men and 0-21% women. Besides, more than 50% men (i.e., male bias) were mentioned in 92.5% of articles, less than 50% men (i.e., female bias) were mentioned in 3.1% of articles and gender equality was observed in 4.4% of articles. If we restricted mentioned people to those who were born after 1960, we observed 50-100% men in ¾ of articles and 100% men in ½ of articles. Considering the percentage of articles with (near to) equal number of genders, a male bias was observed in approx. 72% of articles, gender equality was observed in almost 7% of articles and a female bias was observed in approx. 22% of articles. Thus, we can conclude that elimination of people, who were born before 1960, leads to a less skewed gender inequality towards male bias, meaning that the percentage of articles with male bias decreases. However, male bias remains in approx. 72% of articles.

Analyses of the articles with female title and articles of professions with female majority in the labor market reveals significantly lower ratios of mentioned men. Nevertheless, even for groups where the difference of ratios was significant, a male bias was observed in the larger part of articles from these profession groups.

The data and code are available online: https://github.com/gesiscss/Wikipedia-Language-Olga-master/ .